\documentclass[a4paper,fleqn,usenatbib]{mnras}
\usepackage[T1]{fontenc}
\usepackage{ae,aecompl}

\usepackage{graphicx}	
\usepackage{amsmath}	
\usepackage{amssymb}	
\usepackage{subfig}
\usepackage{scalefnt}
\usepackage{float}
\usepackage[portuges]{babel}
\usepackage[utf8]{inputenc}
\usepackage{scalerel}
\usepackage{tikz}
\usetikzlibrary{svg.path}

\definecolor{orcidlogocol}{HTML}{A6CE39}
\tikzset{
  orcidlogo/.pic={
    \fill[orcidlogocol] svg{M256,128c0,70.7-57.3,128-128,128C57.3,256,0,198.7,0,128C0,57.3,57.3,0,128,0C198.7,0,256,57.3,256,128z};
    \fill[white] svg{M86.3,186.2H70.9V79.1h15.4v48.4V186.2z}
                 svg{M108.9,79.1h41.6c39.6,0,57,28.3,57,53.6c0,27.5-21.5,53.6-56.8,53.6h-41.8V79.1z M124.3,172.4h24.5c34.9,0,42.9-26.5,42.9-39.7c0-21.5-13.7-39.7-43.7-39.7h-23.7V172.4z}
                 svg{M88.7,56.8c0,5.5-4.5,10.1-10.1,10.1c-5.6,0-10.1-4.6-10.1-10.1c0-5.6,4.5-10.1,10.1-10.1C84.2,46.7,88.7,51.3,88.7,56.8z};
  }
}

\newcommand\orcidicon[1]{\href{https://orcid.org/#1}{\mbox{\scalerel*{
\begin{tikzpicture}[yscale=-1,transform shape]
\pic{orcidlogo};
\end{tikzpicture}
}{|}}}}

\usepackage{hyperref} 

\title[Dynamical Environment and Surface Characteristics of Asteroid (16) Psyche]
{Dynamical Environment and Surface Characteristics of Asteroid (16) Psyche}

\author[T. S. Moura et al.]
 {T. S. Moura$^{1}$\thanks{E-mail: santos.moura@unesp.br}\orcidicon{0000-0002-3991-8738}\,
  O. C. Winter$^{1}$\thanks{E-mail: othon.winter@unesp.br}\orcidicon{0000-0002-4901-3289}\,
  A. Amarante$^{2}$\thanks{E-mail: andre.amarante@ifsp.edu.br}\orcidicon{0000-0002-9448-141X}\,
  R. Sfair$^{1}$\thanks{E-mail: rafael.sfair@unesp.br}\orcidicon{0000-0002-4939-013X}\,
  \newauthor
  G. Borderes-Motta$^{3}$\thanks{E-mail: gabriel.borderes@uc3m.es}\orcidicon{0000-0002-4680-8414}\,
  G. Valvano$^{1}$\thanks{E-mail: giulia.valvano@unesp.br}\orcidicon{0000-0002-7905-1788}\
  \\
  $^{1}$ Grupo de Din\^amica Orbital e Planetologia, S\~ao Paulo State University - UNESP, Guaratinguet\'{a}, CEP 12516-410, 
  S\~{a}o Paulo, Brazil\\
$^{2}$ Laboratório Maxwell, Instituto Federal de Educação, Ciência e Tecnologia de São Paulo - IFSP, Cubat\~ao, CEP 11533-160, S\~{a}o Paulo, Brazil\\
$^{3}$ Bioengineering and Aerospace Engineering Department, Universidad Carlos III de Madrid, Leganés, 28911, Madrid, Spain}

\date{Accepted XXX. Received YYY; in original form ZZZ}

\pubyear{2019}

\begin{document}
\label{firstpage}
\pagerange{\pageref{firstpage}--\pageref{lastpage}}
\maketitle

\begin{abstract}
Radar observations show that (16) Psyche is one of the largest and most massive asteroids of the M-class located in the main belt, with a diameter of approximately $230$ km. This fact
makes Psyche a unique object since observations indicated an iron-nickel composition. It is believed that this body may be what was left of a metal core of an early planet that would 
have been fragmented over millions of years due to violent collisions. In this work we study a variety of dynamical aspects related to the surface, as well as, the environment around  this asteroid. We use computational tools to explore the gravitational field generated by this body, assuming constant values for its density and rotation period. We then determine a 
set of physical and dynamical characteristics over its entire surface. The results include the geometric altitude, geopotential altitude, tilt, slope, among others. We also explore the  neighborhood around the asteroid (16) Psyche, so that the location and linear stability of the equilibrium points were found. We found four external equilibrium points, two of them linearly 
stable. We confirmed the stability of these points by performing numerical simulations of massless particles around the asteroid, which also showed an 
asymmetry in the size of the stable regions. In addition, we integrate a cloud of particles in the vicinity of (16) Psyche in order to verify in which regions of its surface the particles 
are most likely to collide.

\end{abstract}


\section{Introduction}

Since asteroids are objects remaining from the early Solar system, we are interested in their composition and internal structure. Consequently, over the years, the agencies NASA, ESA and JAXA 
have studied these objects in their space missions, with the purpose of investigating how the development of the Solar System embryo occurred, as well as the transformations in the  course of 
its evolution. Some missions were designed to obtain information on the physical characteristics of these objects, while others were able to observe asteroids while traveling to other
planetary destinations. For example, the Galileo spacecraft \citep{Galileo1, Galileo2} along its path to the planet Jupiter was the first to observe asteroids. During those encounters, 
high-resolution images of the asteroid (951) Gaspra were obtained in October 1991 and (243) Ida in August 1993. In 1996 NASA launched the NEAR-Shoemaker mission \citep{NShoemaker1, NShoemaker2},
that in addition to orbiting the asteroid (433) Eros also landed on its surface, and in the course of its journey has captured hundreds of images of the asteroid (253) Mathilde. 
The Hayabusa spacecraft \citep{Hayabusa1a, Hayabusa1b}, launched in May 2003 by the Japanese Space Agency, imaged asteroid (25143) Itokawa closely, and for the first time collected material 
from an asteroid returning to Earth for analysis. There is also the Hayabusa 2 mission \citep{Hayabusa2a, Hayabusa2b}, launched in 2014, whose spacecraft has already reached the target, the 
asteroid (162173) Ryugu. This mission has the objective of once again collecting material samples from the object and sending them back to Earth for analysis. 
The European Space Agency started the Rosetta space mission in 2004 \citep{Rosetta1, Rosetta2} to conduct a detailed study of comet 67P/Churyumov-Gerasimenko, and during the course made 
approximations and collected images of the asteroids: (2867) Steins and (21) Lutetia, in 2008 and 2010, respectively. NASA's OSIRIS-Rex mission \citep{Osiris1, Osiris2}, launched in 2016, aims 
to bring samples from asteroid (101955) Bennu. 

Another audacious space mission designed for NASA is the Psyche, scheduled for launch in 2022. The target is asteroid (16) Psyche, the largest asteroid located in the main belt belonging to class M, 
analogous to meteorites composed primarily of iron-nickel (Fe-Ni). An early interpretation of class M asteroids is that its nuclei may originate from violent collisions with planetesimals, 
billions of years ago. It is believed that (16) Psyche is the exposed core of a primitive planet whose composition is Fe-Ni, characterizing it as the unique metallic asteroid located between 
Mars and Jupiter. The main objectives of the space mission are to verify if (16) Psyche really is a nucleus, or if it is only non-molten material, to characterize its topography and to 
estimate the ages of the regions composing its surface \footnote{Website: https://www.jpl.nasa.gov/missions/psyche/}.

Considering that it is a relatively large asteroid with some peculiarities, besides being the target of a future space mission, in this work we will investigate the dynamical environment and surface 
characteristics of (16) Psyche.

Therefore, a study of the gravitational field near these objects is essential. Such a study is complex and can provide dynamic events, such as 
ejection and capture of orbits. Many studies have been conducted to enrich the investigation of the gravitational environment of irregularly shaped bodies.

Considering that the gravitational field induced by small irregular bodies, such as asteroids, is not something simple to calculate, there is a need for increasingly refined methods to assist 
in this complex task. As a consequence of the highly irregular shape of these bodies, the use of the Legendre Polynomials method to calculate the gravitational potential originates some 
discrepancies in certain points. Several strategies have been developed to solve this problem. Triaxial ellipsoids \citep{Scheeres1994}, a massive straight line segment \citep{Riaguas1999} and a 
solid circular ring \citep{Broucke2005} are some approximations used to compute the gravitational field of asteroids. There is also the method of polyhedra created by 
\citet{Werner1994} and \citet{WernerScheeres1996}, where the shape of small irregular bodies is characterized by a polyhedron with constant density, which computes the intensity of the 
gravitational field exerted by these objects. The polyhedra method is more accurate than the methods of the Legendre Polynomials and the use of geometric objects. Thus, several studies have 
shaped the asteroid format using homogeneous polyhedra and applied this method to evaluate the gravitational potential around these bodies. Among the applications is the investigation of the 
dynamic environment and the evolution of orbits around (4179) Toutatis \citep{Scheeres1998}, (433) Eros \citep{Scheeres2000,Chanut2014} and (25143) Itokawa \citep{Scheeres2006}.

In the current work, we adopted a program \citep{Tsoulis2012} that provides the gravitational potential calculated by the polyhedra method, as well as its first-order derivatives, 
with application to (16) Psyche. In order to explore the geophysical environment around this object, we take as reference three works. The first, \citet{Scheeres2012}, presents results 
of the geopotential, surface accelerations and surface slopes on the surface of the asteroids (433) Eros and (25143) Itokawa. The second, \citet{Scheeres2015}, consider the shape of 
small asteroids to be spheres and discusses altitude as a function of slope. Besides,  \citet{Scheeres2016}, in which a series of geophysical calculations were carried out to 
understand the surface of the asteroid (101955) Bennu. These calculations employ the geopotential and its derivatives in different scenarios that will aid in the mapping of the (16) 
Psyche surface. \citet{Scheeres2016} found the surface accelerations, the surface slope, the geometric and geopotential topography, tilt, among other features, and discussed how the 
alteration of the density parameter could influence the geophysical environment of the asteroid (101955) Bennu.

Finally, computing the gravitational potential by the polyhedra method, we can locate the equilibrium points in the gravitational field of small irregular bodies and analyze its linear 
stability, so that the results are more expressive concerning the degree of precision. \citet{Wang2014} demonstrated that the amount of equilibrium points in the gravity field 
of irregular bodies, comprising asteroids and comets, is not fixed, in addition to the existence of at least one point. Since linear stability affects the environment close to the equilibrium 
points, its result can contribute to the good performance of missions with space probes whose objective is to study asteroids, for example. Thus, the investigation of stability and its 
influence in the region near the point of equilibrium can provide support in the construction of a database that would select the best strategy of orbits of reconnaissance around asteroids. 
If a particular parameter of the body varies, the location and stability of the equilibrium points may also change \citep{YuBaoyin2012, Hirabayashi2014, Jiang2015, Jiang2015b}. 
The parameter can be density, rotation speed, format, etc. \citet{Scheeres2016} when studying the dynamic environment of (101955) Bennu also analyzed the behavior of the equilibrium points 
when the density was changed. In this paper, in addition to discussing the characteristics of the geophysical environment of (16) Psyche, we will also present a study of the equilibrium points 
through density change, and for this, we will use the extreme cases since it is a parameter still uncertain \citep{Shepard2017, Drummond2018, Viikinkoski2018}. 

Furthermore, we used a mass concentrations (MASCONS) model \citep{Geissler1996} with the intention of integrating orbits around (16) Psyche. Since the polyhedra method requires several 
calculations to evaluate the potential each time step, the mascons model in addition to maintaining a good accuracy, reduces considerably the simulation time. We performed simulations using 
an adapted version of the Mercury package \citep{chambers1999} that includes the mascons model, considering as initial conditions a ring of massless particles around (16) Psyche. The goal is to 
verify the number of particles that survived until the end of the simulation and those that collided. From the survivors, we can extract information on how long they remain in 
stable regions near the asteroid, in addition to the size of those locations. From the particles that have collided, we will have knowledge of the regions on the surface of 
(16) Psyche that are more favorable to the collision.

The paper consists of the following sections. We discussed in Section \ref{shape} about the shape model and the general characteristics of (16) Psyche. In Section \ref{grav_geop}, we present 
the polyhedra model for computing the gravitational potential and geopotential. The geophysical environment of (16) Psyche is investigated, and the results are presented and discussed 
in Section \ref{results}. The next Section investigates the environment in the vicinity of (16) Psyche through the location and stability of the equilibrium points, as well as an 
analysis of the behavior of massless particles initially arranged around the body and integrated over a given time. In Section \ref{collision}, we present a statistical study of the 
distribution of the impacts of massless particles across the surface of (16) Psyche. Then, we present the final comments in Section \ref{conclusions}.


\section{Shape model and physical properties}
\label{shape}

The analysis of (16) Psyche radar observations confirmed that this object has a much larger albedo than the other main-belt asteroids, leading to the conclusion that it is a metallic-core 
\citep{Shepard2008}. This fact was already exposed in the first radar observations \citep{Ostro1985}. The orbital semi-major axis, inclination, and eccentricity of (16) Psyche are 2.9277 au, 
$3.0950^{\circ}$ and 0.1339, respectively\footnote{Website: https://ssd.jpl.nasa.gov/sbdb.cgi}.

The (16) Psyche shape model constructed by \citet{Shepard2017} was generated by the combination of 18 radar images and 6 continuous wave runs from 2015 with a further 16 continuous wave 
runs from 2005 and 6 adaptative-optics images. This model provided maximum dimensions of $279\times232\times190$ km $(\pm10\%)$. 

\begin{figure}
\begin{center}
\includegraphics*[width=\columnwidth]{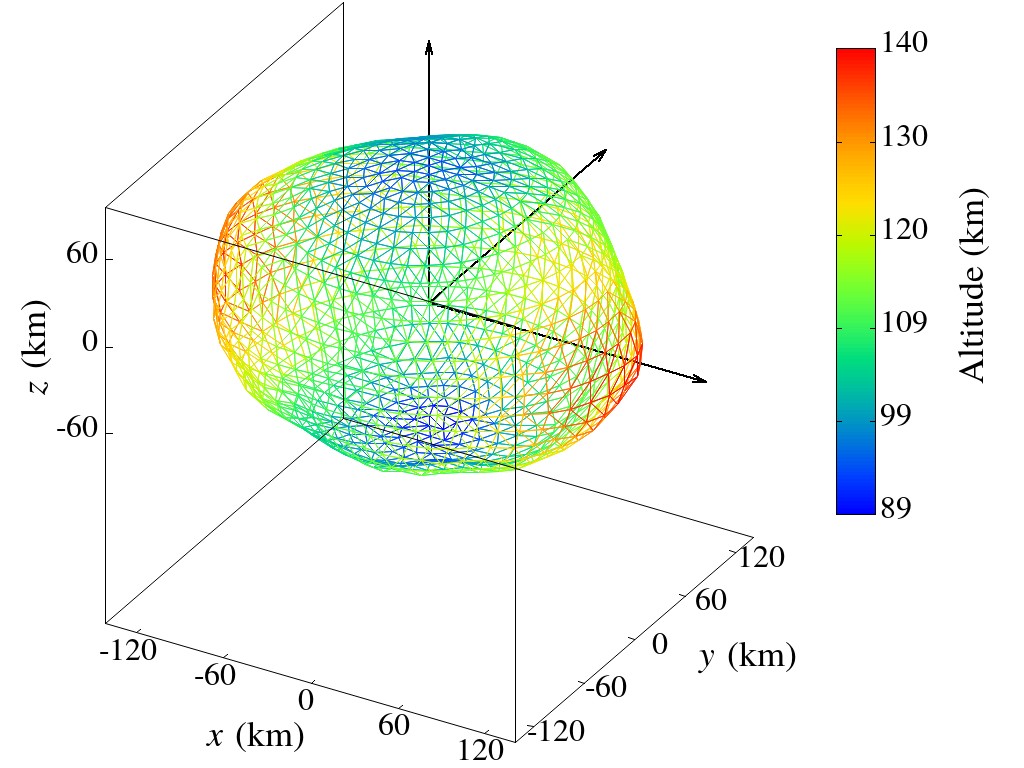}
\end{center}
\caption{Three-dimensional view of the polyhedron shape model of asteroid (16) Psyche. The shape was built with 1148 vertices and 2292 faces and the rectangular color bar provides 
the distance computed from the centre of mass of the object, in km.}
\label{fig: shape}
\end{figure}

For completeness, we present the three-dimensional shape of  the asteroid (16) Psyche. However, a more detailed study can be found in \citet{Shepard2017}. A polyhedron with 1148 vertices and 2292 triangular 
faces covers the irregular surface and  delimits the shape of the asteroid. Fig. \ref{fig: shape} illustrates the three-dimensional polyhedron shape model of (16) Psyche. For the representation of this 
shape model, the origin is  in the centre of mass of the asteroid and the axes $x$, $y$ and $z$ are aligned with the main axes of inertia. We can note that the distance from the centre of mass to the ends 
of the equatorial region is 50$\%$ greater with respect to the same distance from the poles. It is evident from the three-dimensional shape  that the surface of (16) Psyche has several irregularities, 
which may  represent craters of about 10 km depth. Moreover, we can see how the polar regions are flatter than the equatorial region, which makes this object roughly ellipsoidal.

The Infrared Astronomical Satellite (IRAS) mission \citep{Tedesco2002} provides one of the most cited estimates for (16) Psyche size, which is concerning the effective diameter 
(diameter of a sphere whose volume is equivalent to that of the object) $D_{eff} = 253 \pm 4$ km. However, \citet{Shepard2008} observed (16) Psyche in 2005 and the analysis of 8 continuous 
wave runs and 3 delay-Doppler images provided a $D_{eff} = 186 \pm 30$ km, a reasonably lower estimate than that of the IRAS. Although the mass of (16) Psyche has been the subject of 
several estimates given by autonomous methods, there is a concordance of this value that fits in $\sim 12 \times 10^{-12}$ solar mass value \citep{Carry2012}. Following the work of Carry (2012), 
\citet{Shepard2017} adopted the value of $2.72\pm0.75\times10^{19}$ kg as nominal mass. 
Since the density of a body is dependent on the size used, the density of $\rho=3.2$ g cm$^{-3}$ is for the size given by IRAS, while using the model provided by Database of Asteroids Models 
from Inversion Techniques (DAMIT), with $D_{eff} = 190$ km, the density found was $\rho=7.6$ g cm$^{-3}$, material composed of Fe-Ni practically pure and without porosity 
\citep{Shepard2017}.

In addition, \citet{Drummond2018} analyzed a set of adaptive-optics images that resulted in the shape model of (16) Psyche via triaxial ellipsoids of dimensions 
$(a,b,c) = (137, 115.5, 88)$ km. \citet{Drummond2018} also found a $D_{eff} = 223 \pm 7$ km and a new mass value for (16) Psyche, $2.43\pm0.35\times10^{19}$ kg, which led to 
a density estimate of $\rho=4.16\pm0.64$ g cm$^{-3}$. For this density estimation and considering that (16) Psyche is composed of Fe-Ni, we have an exposed, disrupted and reassembled 
nucleus of a planetesimal similar to Vesta, resulting from a macro-porosity of 47$\%$.

Through observations of high-resolution angular images of (16) Psyche from ESO LT/SPHERE/ZIMPOL, \citet{Viikinkoski2018} proposed to refine the three-dimensional shape model of this 
asteroid and, consequently, its density. They analyzed 206 optical lightcurves, 2 stellar occultations, and 38 disk-resolved images. The results indicated a $D_{eff} = 226 \pm 5$ km and a 
mass of $2.41 \pm 0.32\times10^{19}$ kg, which combined generated a new bulk density estimate for (16) Psyche, $\rho=3.99\pm0.26$ g cm$^{-3}$.

The (16) Psyche shape constructed by \citet{Shepard2017} resulted in a $D_{eff} = 226\pm23$ km, a density of $\rho=4.5\pm1.4$ g cm$^{-3}$ and a radar albedo around $0.37$, equivalent to a 
material 40$\%$ porous of metallic composition (Fe-Ni), type regolith. For all the calculations presented in this paper, we adopted the value of the nominal density given by 
\citet{Shepard2017}, as well as the rotation period of 4.195948 h. However, due to the fact that there is some uncertainty in the density value of (16) Psyche, we will make a comparison of 
how its topographic characteristics behave in the extreme values of density (for $\rho=3.1$ g cm$^{-3}$ and $\rho=7.6$ g cm$^{-3}$), in Section \ref{results}.
We will also analyze how the value of this parameter affects the position of the equilibrium points and their linear stability (Section \ref{nearby environment}).
We emphasize that by altering the density of (16) Psyche in Sections \ref{results} and \ref{nearby environment}, we obtain a new volume since we preserve the mass of the body.
The values of $D_{eff}$ computed for the extreme cases of density were 256.31 km and 190.08 km, respectively. Results consistent with those estimated by IRAS and DAMIT observations.

We apply an algorithm \citep{Mirtich1996} that assumes the shape model of (16) Psyche with uniform density and we locate a fixed coordinate system in the body in which the 
origin was transferred to centre of mass of the asteroid. In addition, the $x$, $y$ and $z$ axes were aligned to the principal axes of the smallest, intermediate and largest moments 
of inertia, respectively. We also adopted that (16) Psyche has a uniform rotation on the axis of largest moment of inertia. We obtained a mass of $M = 2.73331\times10^{19}$ kg and a 
volume of $6.07403\times10^6$ km$^3$, values similar to those found by \citet{Shepard2017}. Below are the values, normalized by the mass of (16) Psyche, of the principal moments of inertia:
\begin{equation}
I_{xx}/M = 4.4253\times 10^{3}\ {\rm km}^{2},
\end{equation}
\begin{equation}
I_{yy}/M = 5.4010\times 10^{3}\ {\rm km}^{2},
\end{equation}
\begin{equation}
I_{zz}/M = 6.2298\times 10^{3}\ {\rm km}^{2}.
\end{equation}

We also compute the second degree and order gravity coefficients whose expressions are linked to the principal moments of inertia \citep{Hu2004}:
\begin{equation}
 C_{20}=-\frac{1}{2}(2I_{zz}-I_{xx}-I_{yy})
\label{eq:coefficient C20}  
\end{equation}
\begin{equation}
C_{22}=\frac{1}{4}(I_{yy}-I_{xx}).
\label{eq:coefficient C22}  
\end{equation}
The coefficients $C_{nm}$ express the irregular shape of the mass distribution of a body. It is common to separate the coefficients $C_{nm}$ with index $m$ equal to zero and define 
$J_n =-C_{n0}$, and are called zonal coefficients because they divide the sphere into zones. For the coefficients $C_{nm}$ ($m\geq1$) when $m = n$ are called sector coefficients, they 
divide the sphere into sectors. Then, using the equations (\ref{eq:coefficient C20}) and (\ref{eq:coefficient C22}) we find the gravitational coefficients of (16) Psyche 
(normalized by body mass and radius), in units of distance squared:
\begin{equation}
 C_{20}=-0.1028
 \label{eq:coefficient C20 of Psyche}  
\end{equation}
\begin{equation}
 C_{22}=0.1904.
 \label{eq:coefficient C22 of Psyche}  
\end{equation}
Note that because $I_{xx} \leq I_{yy} \leq I_{zz}$ the gravitational coefficients will have signals $C_{20}\leq 0$ and $C_{22}\geq 0$. We can say that the coefficients $C_{20}$ ($J_2$) and 
$C_{22}$ are the most important and represent the flattening and ellipticity of the body, respectively. For comparison purposes, the values of the gravitational coefficients of the 
asteroid (216) Kleopatra are $C_{20}=-0.6364$ e $ C_{22}=0.3128$, as described in \citet{Chanut2015}.
Note that for (216) Kleopatra to be a highly flattened and elliptical body the values of the coefficients $C_{20}$ and $C_{22}$ must be much larger, in modulus, than the coefficients of 
(16) Psyche.


\section{Gravitational field and geopotential}
\label{grav_geop}

According to \citet{Scheeres2016}, the geopotential expresses the amount of energy flowing on the surface and within a body, and its expression adds contributions of the gravitational 
potential and uniform rotation of that object. When we take into account the motion of a particle in relation to the main body, we can compute the velocity of the particle and the 
value of the geopotential in a given position. The union of the relative kinetic energy with the geopotential results in a conserved quantity, as we shall see further on. Furthermore, 
among the various applications of the geopotential, it is possible to compute the force intensity exerted on the particle according to its location in the body-fixed frame.

Let us consider (16) Psyche rotating with a velocity $\omega$ and being orbited by a massless  particle. It is worth remembering that the shape model that represents the asteroid is 
centred at the centre of mass and aligned with the principal axes of inertia. Then, the expression of the geopotential reduces to \citep{Scheeres2016}:
\begin{equation}
V(\pmb r) = -\frac{1}{2} \omega ^{2}(x^2+y^2) \pmb{+} U(\pmb r),
\label{eq:geopotential}
\end{equation}
where $\pmb r=(x,y,z)$ represents the position of the particle in the body-fixed frame relative to the centre of mass of the body and $U(\pmb r)$ describes the gravitational potential energy.

Using a polyhedra method in the form of a summation, we can compute the gravitational potential as follows \citep{WernerScheeres1996}:
\begin{equation}
 U=\pmb{-}\frac{1}{2}G\rho \sum_{e\in edges}{\pmb r_{\rm e}} \cdotp {\pmb E_{\rm e}} \cdotp {\pmb r_{\rm e}} \cdotp L_{\rm e} \pmb{+} 
 \frac{1}{2}G\rho \sum_{f\in faces}{\pmb r_{\rm f}} \cdotp {\pmb F_{\rm f}} \cdotp {\pmb r_{\rm f}} \cdotp \omega_{\rm f},
 \label{eq: potential}  
\end{equation}
where $G=6.67428\times 10^{-11}\ {\rm m}^{3}\ {\rm kg}^{-1}\ {\rm s}^{-2}$ is the gravitational constant, $\rho$ is the nominal density of the asteroid shape model, 
${\pmb r_{\rm e}}$ and ${\pmb r_{\rm f}}$ are vectors with origin at the field point to any point in the edge $e$ plane and in the face $f$ plane, respectively; 
${\pmb E_{\rm e}}$ and ${\pmb F_{\rm f}}$ are dyads expressed as a function of the two face- and edge- normal vectors related to a given edge and face, and represent the tensors of the edges 
and faces with respect to the field point, respectively; $L_{\rm e}$ is the integration factor, and $\omega_{\rm f}$ is the signed angle viewed from the field point.

The attraction of a body represented by a three-dimensional polyhedron is given by differentiating the potential described by equation (\ref{eq: potential}) \citep{WernerScheeres1996}:
\begin{equation}
 \nabla U=-G\rho \sum_{e\in edges}{\pmb E_{\rm e}} \cdotp {\pmb r_{\rm e}} \cdotp L_{\rm e} + 
 G\rho \sum_{f\in faces}{\pmb F_{\rm f}} \cdotp {\pmb r_{\rm f}} \cdotp \omega_{\rm f}.
 \label{eq: attraction}  
\end{equation}


\subsection{Equation of motion}
\label{jacobi_constant}

The motion of a particle near an asteroid may be influenced by perturbations of the Sun, mainly by gravity and the radiation pressure force. Such perturbations  are sometimes important  because they assist 
in the investigation of the dynamic environment around these small bodies. The increase in the distance of the particle from the asteroid intensifies the action of solar gravity.
The radiation pressure, in addition to depending on the distance and direction of the Sun with respect to the asteroid, also takes into account the ratio of area to mass of the particle. 
Consequently, as the distance between asteroid and particle increases, this perturbation produces a more relevant contribution.

Then, the movement of a particle orbiting (16) Psyche, for example, can be significantly influenced by a set of perturbative forces, such as the effects of solar gravity, radiation pressure  and the 
irregular gravitational potential of the central body. However, in many situations, a single force may be strong enough to end up being dominant in the system. In this way, we can 
neglect the weaker forces without loss of information on the resulting orbital dynamics. \citet{Hamilton1996} presented an analytical and numerical study of the behavior of such  perturbative forces 
in the dynamics of circumplanetary dust grains. They have considered the most significant effects of solar gravity from the so-called tide term, neglecting the higher-order 
solar terms, since the intensity of these terms, is up to 100 times weaker. For the radiation pressure, they considered a dust grain in the form of a sphere with uniform density and 
scattering properties, in addition to the absence of variations in the direction and intensity of the radiation pressure force over the trajectory of the dust grain (parallel-ray approximation). 
Furthermore, for a non-spherical mass distribution of a central body, the term $J_2 = -C_{20}$ is usually the dominant one.

\begin{figure}
\begin{center}
\includegraphics*[trim = 0mm 0mm 0mm 22mm, width=\columnwidth]{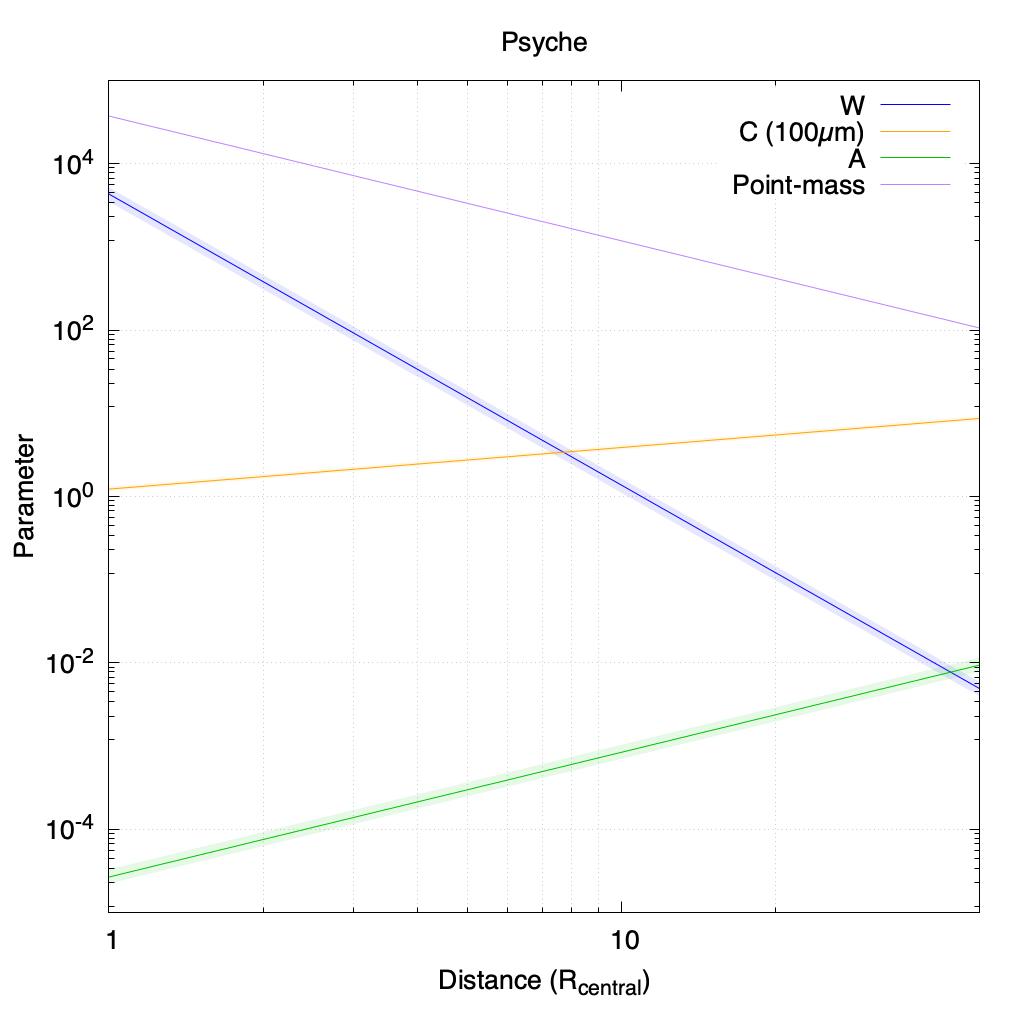}
\end{center}
\caption{Variation of the parameters $A$, $C$ and $W$ representing the intensity of the solar tidal force, the radiation pressure and the oblateness, respectively, as a function of the 
distance from (16) Psyche. For each parameter there is a shaded region indicating the variation between the pericentre and apocentre. $R_{\rm central}=113.2$ km is the equivalent radius of (16) Psyche.}
\label{fig: dimensionless parameters}
\end{figure}

For each perturbative force \citet{Hamilton1996} introduced dimensionless parameters in order to compare the perturbation strengths. The parameters are: solar tidal parameter $A$, radiative 
parameter $C$ and oblateness parameter $W$. The calculation of such parameters is expressed in relation to some physical properties of the central body and the dust particle.
In Fig. \ref{fig: dimensionless parameters} we present the variation of parameters $A$, $C$, and $W$ as a function of distance from (16) Psyche, and for comparison, there is a line 
indicating a scaled strength for the point-mass attraction of the asteroid. For each parameter, there is a shaded region indicating the variation of their values between the pericentre  and apocentre. 
We can note a decay in both oblateness ($W$) and point-mass parameters as distance increases since both effects are $\propto~ 1/r^2$. In the other hand, the radiation pressure ($C$) and the solar tide ($A$) 
are stronger for regions further from the asteroid, where a particle experiences a weaker gravity from the central body, and it is more susceptible to be disturbed by the effects that arise from the Sun.
We restrict our analysis to grains above 100~$\mu$m. At this size range, the gravitational field of (16) Psyche is strong enough to dominate the orbital motion of particles the asteroid, and any 
solar perturbation (no matter if radiation or gravitational) can be safely ignored. For instance, the gravitational parameters ($W$ and point-mass) are at least 
three orders of magnitude larger than the radiation and solar tidal coefficients. Thus, our results cannot be extended to finer dust grains for which the solar radiation force may become relevant. 

Therefore, in this paper, we consider that the gravitational field of (16) Psyche is strong enough to dominate the orbital motion of particles around it.
Consequently, we are neglecting any solar perturbations and also the influence of the gravitational force from other bodies. Since at a distance of twice the equivalent radius of 
(16) Psyche the parameters $C$ and $A$ are at least $10^2$ and $10^6$ times smaller, respectively than the parameter $W$. Thus, the asteroid gravitational force is much stronger than the 
other perturbations.

Then, considering that the rotation of (16) Psyche is uniform about its axis of maximum moment of inertia, the equations of motion of a particle in its vicinity are defined by 
\citep{Scheeres2016}:
\begin{equation}
 \ddot{\pmb r} +2{\pmb \omega}\times\dot{\pmb r} = - \frac{\partial V}{\partial \pmb r},
 \label{eq: equations of motion}
\end{equation}
where $\dot{\pmb r}$ and $\ddot{\pmb r}$ express velocity and acceleration of the particle. The existence of an integral of motion, which arises from the fact that equation 
(\ref{eq: equations of motion}) is time invariant, can be related to energy in the body-fixed frame. The Jacobi constant $J$, as this conserved quantity is called, is computed as 
follows \citep{Scheeres2016}:
\begin{equation}
J = \frac{1}{2}v^2+V(\pmb r),
\label{eq: jacobi}
\end{equation}
where $v$ is the intensity of the velocity vector with respect to the rotating asteroid. 


\section{Topographic features}
\label{results}

The mapping of the geopotential and its derivatives, through the surface of (16) Psyche, can be used to compute some quantities that contribute to the investigation and identification of the 
topographic features of this object. The geopotential, as defined by equation (\ref{eq:geopotential}), is the combination of the gravitational potential added to the centrifugal potential. 
We emphasize that the latter is directly related to the spin rate of the body, which we assume to be rotating uniformly on the axis of the largest moment of inertia.

Thus, we will characterize the surface of (16) Psyche taking into account the gravitational and rotational effects. The contribution of these two effects makes it possible to determine 
the relative energy in any region on the surface of the body. One way to compute this energy is by defining a reference value, called ``sea-level'' height, so that we can measure an 
effective altitude at different locations above that reference. Another more significant way is to connect the geopotential directly to possible dynamic motions across the surface of 
the asteroid. The displacement of a particle on the surface of the body requires a certain velocity. So, we can compute the geopotential, or, more precisely, the relative energy, as a function 
of the kinetic energy or velocity necessary to propel such motion. Thus, we will analyze the geopotential under these two aspects.

In this paper, we first compute the gravitational potential of (16) Psyche on its surface, given by equation (\ref{eq: potential}). As the shape model of 
this object was based on polyhedra with triangular faces, we computed the value of the gravitational potential in the barycentre of each face. And then, we obtained the 
geopotential value on the surface of (16) Psyche, according to equation (\ref{eq:geopotential}). The result allowed us to characterize the surface of this object employing the 
geopotential altitude, potential speed, surface accelerations, and the mapping of slope angle, as we will see from Subsection \ref{geopotential topography} onwards.

In addition, we investigated the topography of (16) Psyche taking into account only geometric aspects, without the influence of its properties, such as gravitational field and rotation 
speed. The analysis of how each triangular face that composes the polyhedral shape model of (16) Psyche behaves in relation to the body geometry is characterized by the geometric altitude and 
tilt angle quantities, defined in the next subsections.


\subsection{Geometric topography}
\label{geometric topography}

\begin{figure*}
\begin{center}
\includegraphics[height=9cm, width=13cm]{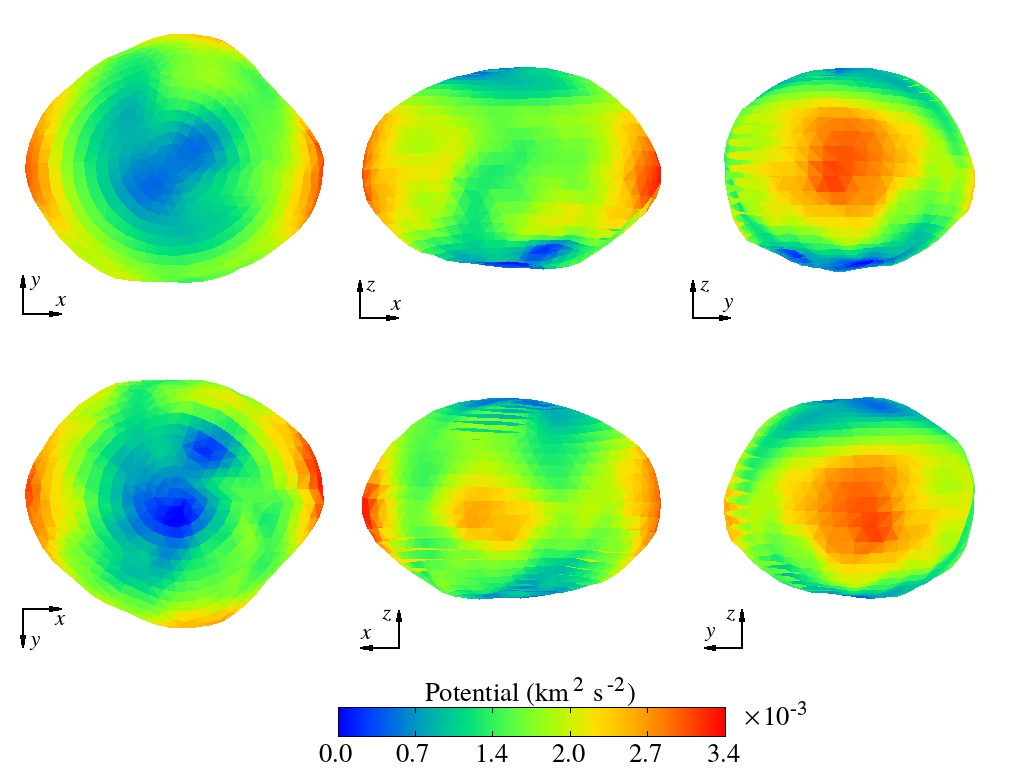}
\end{center}
\caption{Gravitational potential mapped across the surface of (16) Psyche considering the nominal density, under different views, computed in relation to ``sea-level'' value 
($-1.7710\times 10^{-2}$ km$^2$ s$^{-2}$).}
\label{fig:potential}
\end{figure*}

\begin{figure*}
\begin{center}
\includegraphics[height=9cm, width=13cm]{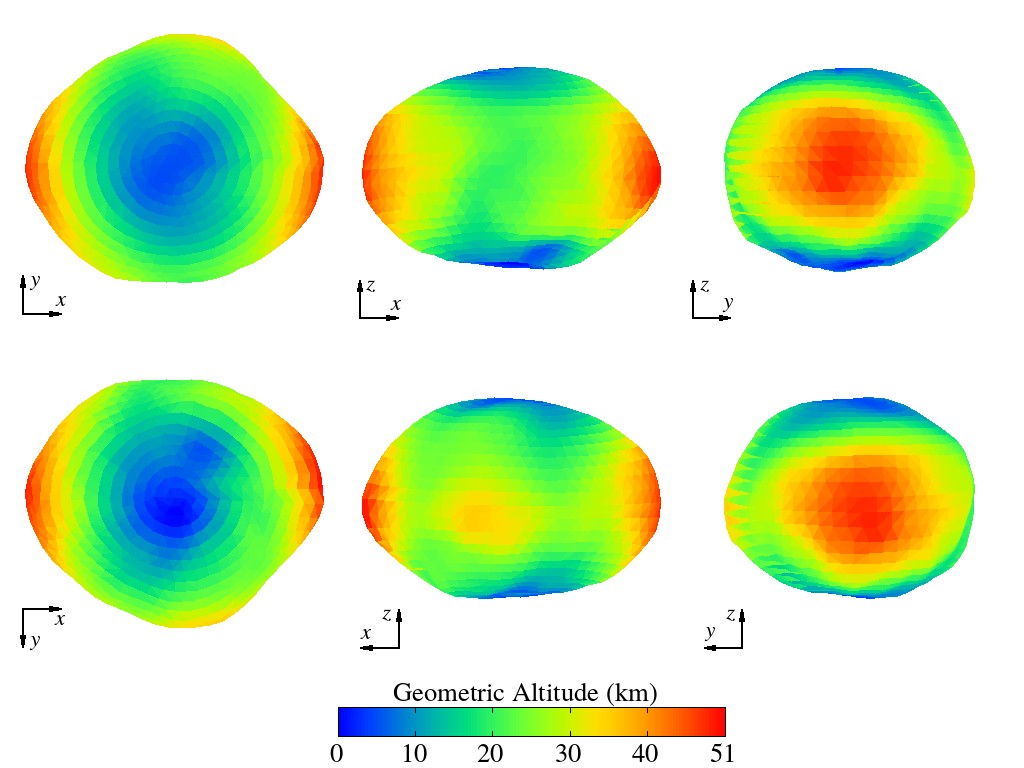}
\end{center}
\caption{Geometric altitude mapped across the surface (16) Psyche considering the nominal density, under different views. Geometric altitude is computed from the lowest magnitude of the 
radius vector on the surface ($89.2366$ km).}
\label{fig:geometric}
\end{figure*}
The definition of \textit{geometric topography} of the body takes into account variations in the radius of the body with respect to a minimum radius, and also the orientation of the surface 
in relation to the vector whose magnitude is the radius of the body \citep{Scheeres2016}. The radius is measured from the origin in the body-fixed coordinate frame to the barycentre of 
each triangular face that composes the asteroid shape model. The first definition represents the geometric altitude while the second gives us the value of an angle known as ``tilt'' 
(Subsection \ref{tilt}).

In general, for planetary bodies, we observe that the variation in altitude depends on the geometry of the surface. Both are intimately connected so that an alteration in the direction of one 
also represents an alteration in the direction of the other. Thus, if the radius of the body increases or decreases according to the region at the surface, the gravitational potential will also 
increase or decrease, and consequently, there will be a change in relative altitude. We then compute the potential energy across the surface of (16) Psyche relative to a 
reference value. This reference value is considered as a ``sea-level'' value, or simply the smallest value computed across the surface \citep{Scheeres2012, Scheeres2015}.
In Fig. \ref{fig:potential} we have the potential energy change on the surface of (16) Psyche, measured in relation to ``sea-level''. While Fig. \ref{fig:geometric} exposes the change in 
radius of the body, measured in relation to a minimum radius, according to the geometry of the surface. We clearly note the previously cited relationship between altitude and potential energy.
We have a total altitude variation of approximately 50.5 km in relation to the minimum radius. Note that the greatest variation occurs at the equator in the region surrounding the 
ends of the object, which can also be seen in Fig. \ref{fig: shape}.

We also obtained the total variation in altitude for the extremes of densities discussed in \citet{Shepard2017} ($\rho=3.1$ g cm$^{-3}$ and $\rho=7.6$ g cm$^{-3}$). We computed a variation of 
57.2 km for the minimum density, and for maximum density, this value decreases to 42.4 km. Although the density difference is relatively large, the variation in geometric altitude does not 
have discrepant values, so the graphs that present these results are similar to that of Fig. \ref{fig:geometric}. Note that because we are preserving the mass of (16) Psyche, we have the 
following relation: as the density increases, the volume decreases, and consequently, the variations in the geometric altitude also decrease.


\subsection{Tilt}
\label{tilt}

\begin{figure*}
\begin{center}
\includegraphics[height=9cm, width=13cm]{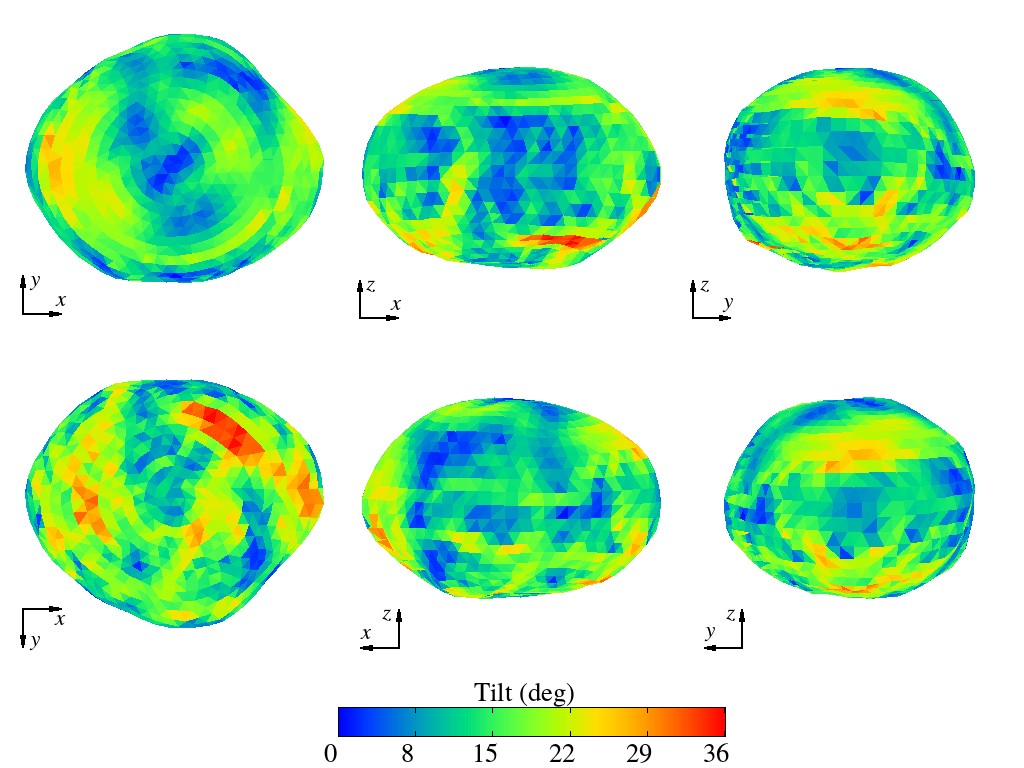}
\end{center}
\caption{Tilt angle mapped across the surface of (16) Psyche considering the nominal density, under different views.}
\label{fig:tilt}
\end{figure*}
As already mentioned at the beginning of Subsection \ref{geometric topography} the definition of geometric topography carries with it the surface orientation, which is 
characterized by the angle tilt. Given a point $\pmb r$ any on the surface of (16) Psyche. The direction of the vector normal to the surface at this point $\pmb r$ establishes the 
orientation of the surface at that location, in the body-fixed frame. Similarly to \citet{Scheeres2016}, we adopt as the defining direction the vector that starts from the body center 
of mass and points to the location $\pmb r$ on the surface of (16) Psyche, that is, the vector whose magnitude is the radius at that surface location. Then, the angle between the normal 
vector the surface and the radius vector is called the tilt angle of the location $\pmb r$.
To compute the tilt angle based on the shape model of (16) Psyche, we consider a point $\pmb r$ on each triangular face and find the normal vector on each face and the vector radius 
according to the location $\pmb r$. This angle helps to understand how each triangular face, which composes the form of (16) Psyche, is oriented in the structure of the body.

Fig. \ref{fig:tilt} presents the mapping of the tilt angle across the surface of (16) Psyche. Observe that the variation of the tilt angle is relatively small, not exceeding $36^{\circ}$. 
In addition, the highest values of this angle lie in a region close to the poles. And if the object in question was a sphere, by definition given, the tilt angle 
would have a constant value equal to zero. Since the (16) Psyche shape resembles a flat sphere, but with irregularities over its entire surface, as seen in Fig. \ref{fig: shape}, this variation 
of the tilt angle given in Fig. \ref{fig:tilt} is adequate.

Therefore, depending on how irregular the asteroid shape is and also the location on its surface, we can have significant variations of the tilt angle.
Note that this is an angle that depends exclusively on the geometry of the body, which will be the same regardless of the change in density.


\subsection{Geopotential topography}
\label{geopotential topography}

\begin{figure*}
\begin{center}
\includegraphics[height=9cm, width=13cm]{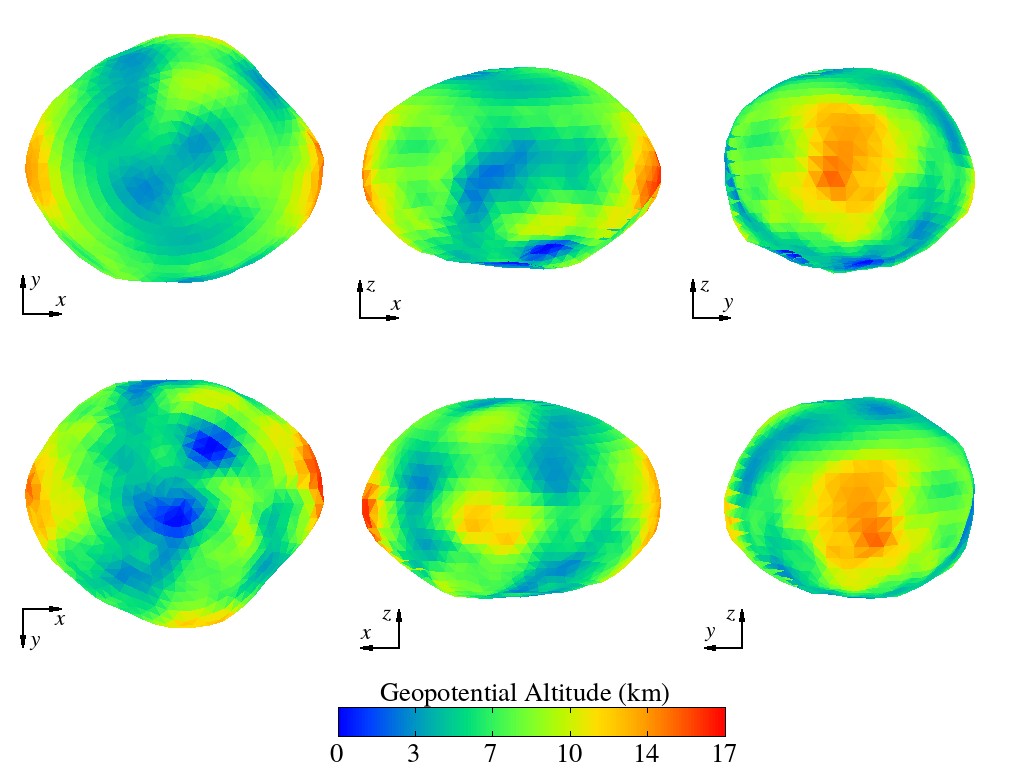}
\end{center}
\caption{Geopotential altitude mapped across the surface of (16) Psyche considering the nominal density, under different views. Geopotential altitude is computed from the lowest geopotential 
value on the surface. The minimum value of the effective altitude on the surface of (16) Psyche is 128.18 km.}
\label{fig:geopotential}
\end{figure*}
Also in \citet{Scheeres2016} we find the definition of \textit{geopotential topography} of the body established by the junction of two factors, they are: modifications in the geopotential of 
the object and how inclined the surface (each triangular face of the asteroid shape model) is in relation to the acceleration vector, determined from the geopotential. The first definition 
represents the geopotential altitude (also specified in \citet{Turcotte2014}) while the second is a mean to find the value of an angle called ``slope'' 
(Subsection \ref{slope}).

As the purpose is to investigate the geopotential altitude across the surface of (16) Psyche, we first map the geopotential on its surface and define the reference value ``sea-level'' to be 
the lowest value computed. \citet{Shepard2017} also computed the geopotential altitude across the surface of (16) Psyche, but in relation to a mean geopotential value.
We map the geopotential again in various locations on the surface of the body, but now in relation to the ``sea-level'' value $(-1.7785\times 10^{-2}$ km$^2$ s$^{-2})$. Then, this new 
result was divided by the value of the local total acceleration of the geopotential in these several locations. This division originated an effective altitude measured in units of length. 
Fig. \ref{fig:geopotential} illustrates the behavior of the geopotential altitude across the surface of the body we are studying. Comparing Figs. \ref{fig:geometric} and \ref{fig:geopotential} we can 
see a certain relation between the geometric changes in the surface and the changes in the geopotential. However, this correspondence usually does not occur on the surface of small bodies whose rotation 
speed is high, due to the great contribution of the rotational component in the geopotential, as shown in the work of \citet{Scheeres2016} for the asteroid (101955) Bennu. Although (16) Psyche 
and (101955) Bennu have a similar rotational period, the radius of the first is about 453 times the radius of the second. In this way, the influence of the rotational effect on the geopotential 
ends up being balanced and even smaller than the gravitational effect. We note from Figs. \ref{fig:potential} and \ref{fig:geopotential}, that the rotational effect interferes more in 
the equatorial region of (16) Psyche.

We know that work is the measure of energy transferred by the application of a force along a displacement and that this force is given by the negative gradient of the geopotential.
Therefore, we can correlate the altitude variation indicated in Fig. \ref{fig:geopotential} with an estimate of the quantity of work demanded in the displacement of a particle on the 
surface of (16) Psyche that travels, for example, from a certain altitude to a higher altitude. In addition, the density value of (16) Psyche is also a parameter that influences the occurrence 
of changes in the effective altitude map.
\begin{figure*}
\begin{center}
\includegraphics[height=9cm, width=13cm]{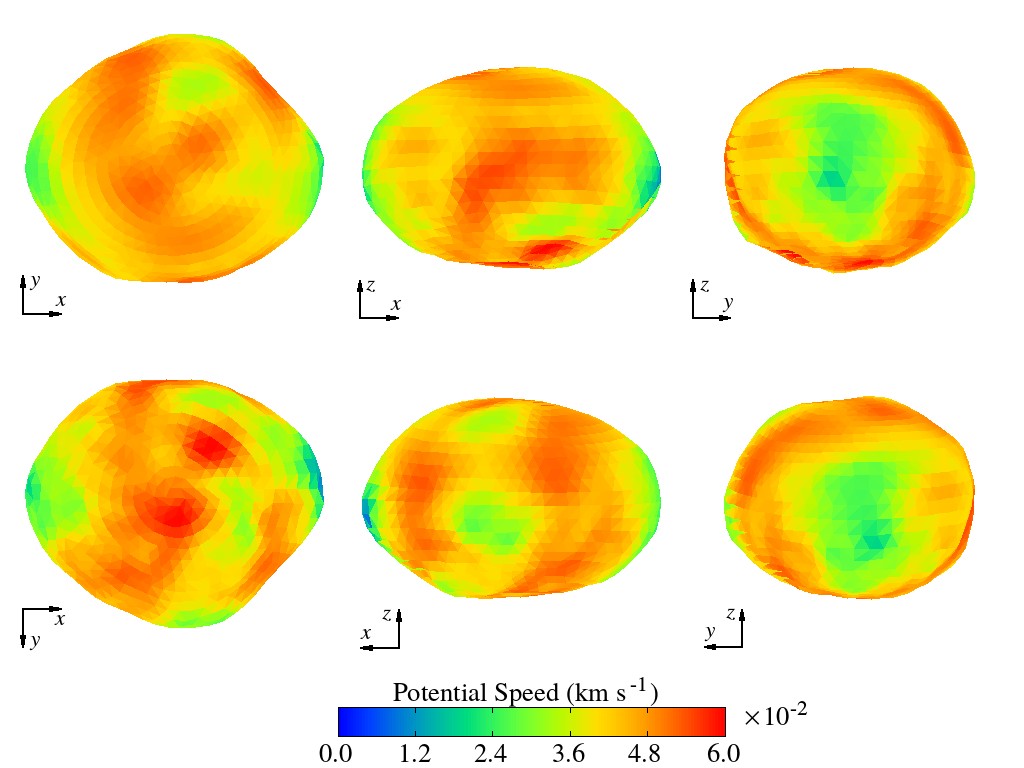}
\end{center}
\caption{Relative Jacobi speeds $\Delta v_j$ evaluated across the surface of (16) Psyche considering the nominal density and $v_j^m=1.7878 \times 10^{-1}$ km s$^{-1}$, under different 
views. The values indicate the velocity that would be achieved by a particle that starts its trajectory from the highest point in the geopotential (at the ends of the equatorial region) 
towards the lowest point (in the pole region).}
\label{fig: view_speeds}
\end{figure*}
Since the total variation in altitude varies as a function of density, the behavior of the geopotential altitude under the minimum, nominal, and maximum density values undergoes total 
variations equivalent to 14.2 km, 16.9 km, and 17.6 km, respectively. Although (16) Psyche is an object of fairly large size, it does not present significant changes in the total variation 
in altitude considering extreme values of density. Thus, the graphs that represent these results are similar to Fig. \ref{fig:geopotential}, with the largest variations occurring in the 
equatorial region.


\subsection{Potential speed}
\label{pot_speed}
The displacement of particles across the surface of (16) Psyche provides a more detailed analysis of the geopotential. Let us suppose a particle at rest at a location $\pmb r' $ on the surface 
of the asteroid. We saw previously in Subsection \ref{jacobi_constant} that the Jacobi integral, expressed in Eq. (\ref{eq: jacobi}), connects the velocity of this particle with the 
geopotential computed at position $\pmb r' $ and the value of its Jacobi constant. By means of this relation and according to the position of the particle in the body-fixed frame, we 
can transform the energy available for the motion of the particle between any two locations in kinetic energy. Then, if the particle located at point $\pmb r' $ moves to another point 
$\pmb r'' $, on the surface of (16) Psyche, the kinetic energy necessary for that purpose can be compared. So, the values of the Jacobi integral evaluated at points $\pmb r' $ and $\pmb r" $ can 
be confronted to find the solution of the following expression:
\begin{equation}
\frac{1}{2}v''^2+V(\pmb r'')=V(\pmb r'),
 \label{eq: jacobi1}
\end{equation}
whose result gives the speed required for a particle to move between two different sites on the surface of the asteroid
\begin{equation}
\label{eq: velocity}
v'' = \sqrt{-2(V(\pmb r'')-V(\pmb r'))}.
\end{equation}
The trajectory of a particle from position $\pmb r' $ to $\pmb r'' $ is only feasible if the result of subtraction $(V(\pmb r'')-V(\pmb r'))$ is negative, otherwise, movement will not be 
allowed. Thus, the difference between the relative energies computed at the positions $\pmb r' $ and $\pmb r'' $ is expressed by kinetic energy, or more precisely in relation to 
the velocity that the particle must have in order to travel between those locations. As the particle moves between the regions on the surface of the asteroid, there will be a loss or a gain 
in speed. Consequently, we can map the geopotential across the surface of (16) Psyche using this relevant concept of velocity, and find the lowest and highest geopotential locations.

\citet{Scheeres2016} defined this concept as ``Jacobi speed'' in the form $v_j=\sqrt{-2V(\pmb r)}$, with evident dependence of the position $\pmb r$ of the particle on the surface of 
the asteroid. Considering that this velocity can be defined at any location on the surface of (16) Psyche, then the maximum value of $v_j$ occurs at the point of lowest geopotential.
So, $v_j^m$ (minimum value among those calculated on the surface of the object) sets the highest point in the geopotential, and relative values of the Jacobi speed mapped on the 
asteroid surface are computed as follows
\begin{equation}
\label{eq: velocity2}
\Delta v_j(\pmb r) = \sqrt{(v_j(\pmb r))^2-(v_j^m)^2}.
\end{equation}
The above relationship establishes the speed gain necessary to boost the trajectory of a particle that leaves the highest point in the geopotential to any location $\pmb r$ on the surface of the 
asteroid. However, another interpretation would be to relate this equation to the velocity that a particle must have at the beginning of its trajectory at point $\pmb r$ toward the highest 
point in the geopotential.

Fig. \ref{fig: view_speeds} maps the quantity $\Delta v_j$ across the surface of (16) Psyche, taking into account the nominal density value. Observe that it is at the extremities of the 
equatorial region of the body where the maximum geopotential value is located since this location comprises the highest point in the geopotential. While the geopotential minimum is
preferably located in the polar regions since this location comprises the lowest point in the geopotential. This analysis is in full accordance with the map of the effective altitude, 
given by Fig. \ref{fig:geopotential}.
We conclude that for (16) Psyche $v_j$ is minimal at the extremes of the equatorial region and maximum at the poles. 
Thus, a particle moving from the equatorial region to one of the poles requires an amount of $\Delta v_j$ to boost its motion on the order of 0.06 km s $^{-1}$. 
This is clearly shown in Fig. \ref{fig: view_speeds}.

Although we do not present the graphs with the results of $\Delta v_j$ across the surface of (16) Psyche for the extreme values of density, we report that the overall magnitude of the 
speeds increases as the density increases. Fact expected since the geopotential depends directly on the density of the object.
Moreover, even for the extreme densities, the behavior of $\Delta v_j$ follows the same pattern as that of Fig. \ref{fig: view_speeds}.
For the low density the total variation of speeds is $4.79\times 10^{-2}$ km s$^{-1}$, $6.01\times 10^{-2}$ km s$^{-1}$ for the nominal and $7.62\times 10^{-2}$ km s$^{-1}$ for the high density, 
relatively large changes.


\subsection{Surface accelerations}
\label{surfaccel}
Let us again consider a particle in position $\pmb r$ on the surface of (16) Psyche. The total acceleration experienced by the particle at this point $\pmb r$ is the junction of the 
gravitational and centrifugal accelerations. We can map the total acceleration to any region in the body-fixed frame. Based on the shape model of (16) Psyche, we locate the coordinates of the 
barycentre of each triangular face and then, we obtain the total acceleration through the negative gradient of the geopotential $-\partial V/ \partial \pmb r$ computed in these points.

\begin{figure*}
\begin{center}
\includegraphics[height=9cm, width=13cm]{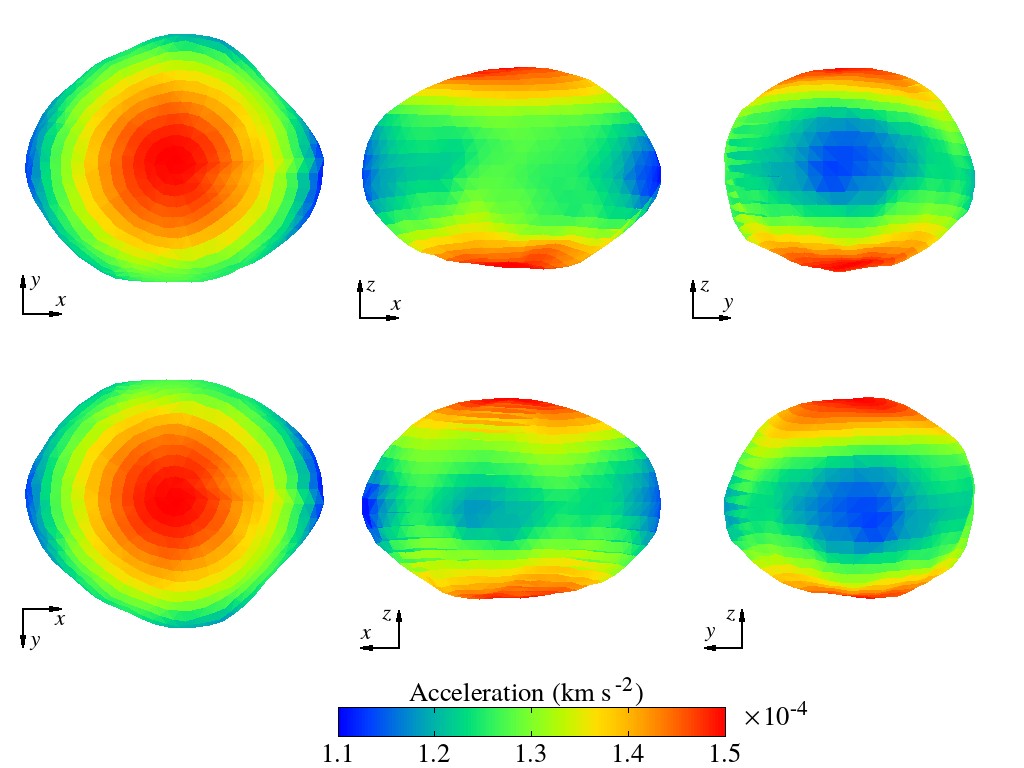}
\end{center}
\caption{Total acceleration mapped across the surface of (16) Psyche considering the nominal density, under different views.}
\label{fig:acceleration}
\end{figure*}

Fig. \ref{fig:acceleration} indicates the values of the total acceleration across the surface of (16) Psyche considering the nominal density. We can observe that the acceleration is stronger 
in the south and north poles of (16) Psyche, while in the equatorial region, especially in the extremities, its influence is minimized. This behavior reflects the competition between the 
terms expressing the gravitational and centrifugal accelerations computed from $-\partial V/ \partial \pmb r$. Although it is evident from Fig. \ref{fig:acceleration}, by the arrangement of 
the colors representing the effect of the total acceleration across the surface of (16) Psyche, the variation between the values comprising the minimum and maximum acceleration is very 
low, less than $4.0\times 10^{-5}$ km s$^{-2}$. In addition, we can decompose these accelerations and analyze the normal components to the local surface and tangential to the local surface.
The accelerations normal to the surface range from 1.027 to 1.451 km s$^{-2}$, while the accelerations tangent to the surface range from 0.003 to 0.486 km s$^{-2}$.

The results for the extreme values of density also exhibit the same behavior as that indicated in Fig. \ref{fig:acceleration}. Specifically, the maximum total acceleration across the 
surface of (16) Psyche is $1.13\times 10^{-4}$ km s$^{-2}$ for the lowest density, $1.45\times 10^{-4}$ km s$^{-2}$ for the nominal and $2.06\times 10^{-4}$ km s$^{-2}$ for the highest 
density. This represents a factor of 1.8 between the maximum magnitudes of the accelerations considering the extreme values of the density. Therefore, the change in density influences 
considerably the intensity of total acceleration experienced by a particle on the surface of the body.

For comparison purposes, the total magnitudes of the accelerations computed on the surface of the asteroid (101955) Bennu are less than $1.00\times 10^{-7}$ km s$^{-2}$ \citep{Scheeres2016}, 
about a thousand times smaller than the magnitudes of Fig. \ref{fig:acceleration}.

\begin{figure*}
\begin{center}
\includegraphics[height=9cm, width=13cm]{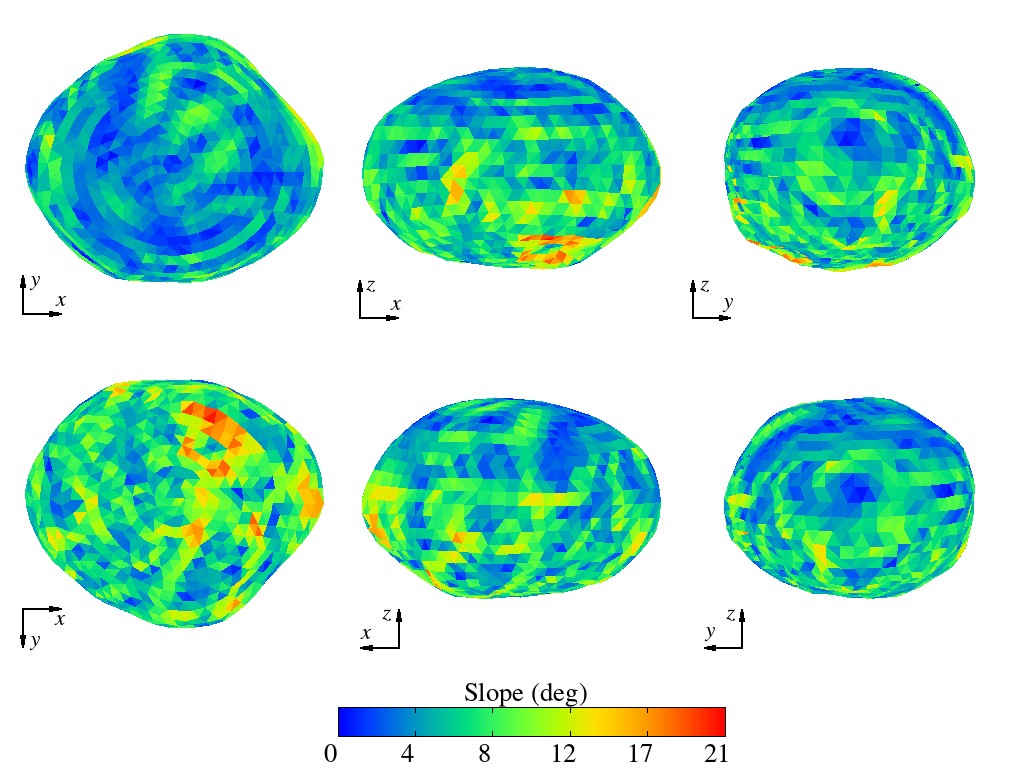}
\end{center}
\caption{Slope angle mapped across the surface of (16) Psyche considering the nominal density, under different views.}
\label{fig:slope}
\end{figure*}

\begin{figure}
\begin{center}
\subfloat{\includegraphics*[trim = 0mm 30mm 0mm 2.5mm, width=\columnwidth]{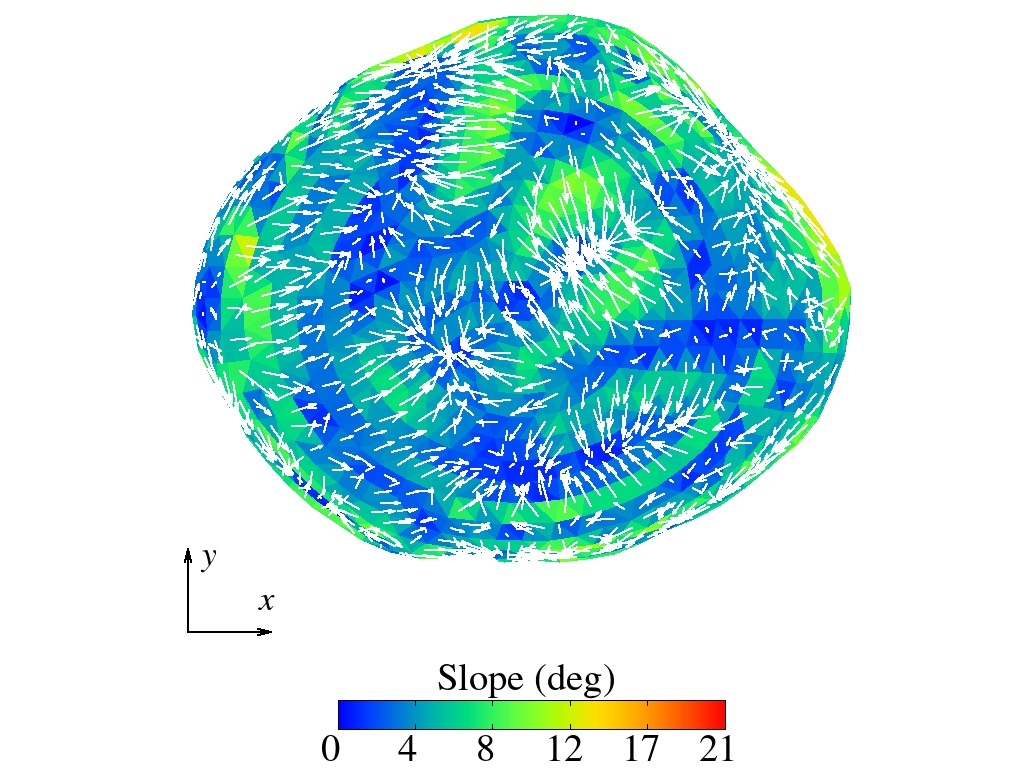}}\\
\subfloat{\includegraphics*[trim = 0mm 30mm 0mm 2.5mm, width=\columnwidth]{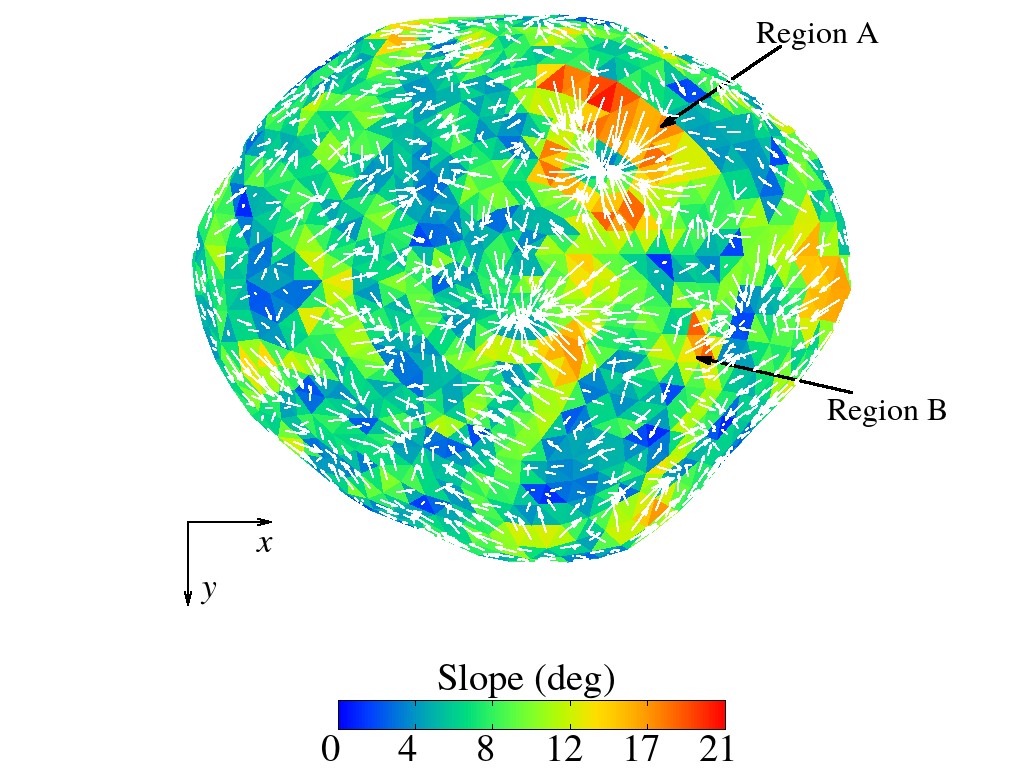}}\\
\subfloat{\includegraphics*[trim = 0mm 0mm 0mm 2.5mm, width=\columnwidth]{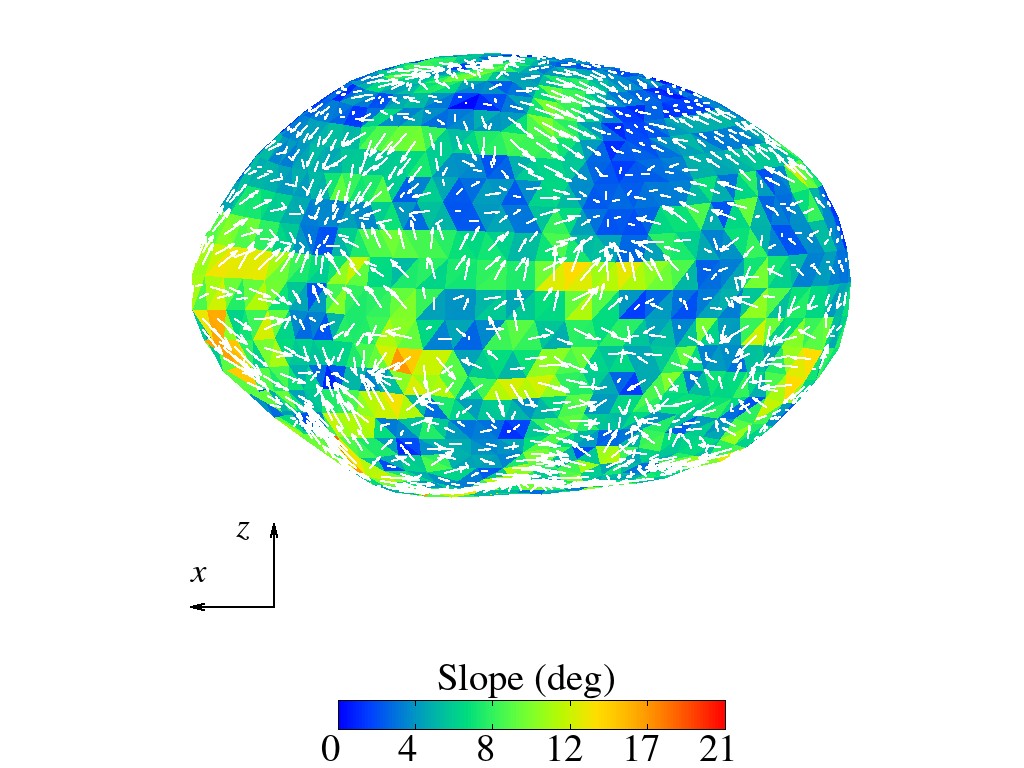}}
\end{center}
\caption{Directions of the tangential acceleration vectors mapped across the surface (16) Psyche considering the nominal density,
under the views of the northern and southern hemispheres and equatorial region, respectively.}
\label{fig: vector slope}
\end{figure}

\subsection{Slope}
\label{slope}
Taking into consideration Subsection \ref{geopotential topography}, the geopotential topography also expresses how the surface is oriented in relation to the acceleration 
vector, computed from the geopotential at a given location on the surface of the body. Consider a location $\pmb r$ on the surface of (16) Psyche, the relative orientation between the 
vector normal to that surface, and the total acceleration vector computed at point $\pmb r$ establish the slope surface \citep{Scheeres2016}.
Then, the slope angle is determined through the supplement of the angle formed by the normal vector to the surface and by the total acceleration vector at point $\pmb r$.
We emphasize that, according to the definition, if these vectors point to opposite directions, the slope angle will be zero; if they are perpendicular, it will be $90^{\circ}$.

Fig. \ref{fig:slope} contains the mapping of the slope angle on the surface of (16) Psyche, considering the nominal density. To generate these maps, we first locate the normal surface 
vector and the total acceleration vector in the barycenter of each triangular face of the three-dimensional shape model of (16) Psyche and then compute the slope angle at each of these points.
Note that the slopes on the surface of (16) Psyche do not exceed $21^{\circ}$, assuming nominal density, as computed in \citet{Shepard2017}.
The distribution of maximum slopes is preferably concentrated in the vicinity of the southern hemisphere and with some peaks at the equator, and the lower slopes generally occur near 
the north hemisphere. The results revealed that about 84$\%$ of the (16) Psyche surface has a slope less than $10^{\circ}$.

Since the surface acceleration is divided into a normal and tangential component, the acceleration vectors tangent to the surface generally point towards regions 
with slopes close to zero. 
In Fig. \ref{fig: vector slope} we present the mapping of the accelation vectors tangent to the surface of (16) Psyche, highlighting the northern and southern hemispheres and equatorial region. 
Note that there is a correlation between the direction of these vectors and the apparent downslope motion of material on almost the entire surface of (16) Psyche.
Consequently, a particle abandoned from rest at a location $\pmb r$ on the surface of Psyche (16) tends to have its movement directed towards 
these low slope regions. In this way, there will be places on the surface of the body that are predisposed to accumulate material \citep{Scheeres2012}.
As an example we identified Regions A and B in the southern hemisphere of the body (reddish region in Fig. \ref{fig: vector slope}), which are places that concentrate the highest slope values, 
and whose direction of the tangential acceleration vectors clearly point to low slope regions located near Regions A and B. Moreover, we clearly observed a significant difference in the behavior of the 
slope angle in the northern and southern hemispheres. 
Since blue delimits slopes close to zero, these regions around the north hemisphere represent stable resting areas for material concentration, while 
regions near the south hemisphere are unstable. And, consequently, in these regions will occur a more intense flow of material. 

Finally, in the equatorial region of Fig. 10 we notice a wide central band with low slopes ranging from $8^{\circ}$ to $10^{\circ}$, with a few locations whose value is slightly higher.
The tangential acceleration vectors in this band point mainly to even lower slope regions, which are located closer to the northern and southern hemispheres. Therefore, we have that the flow of 
loose material on the surface of (16) Psyche will migrate preferentially towards the regions near the northern and southern hemispheres.

However, for bodies much smaller than (16) Psyche, such as asteroids (1999) KW Alpha and (101955) Bennu, although with a similar rotation period as 16 (Psyche), the slopes mapped through the 
surface of these objects exhibit a pattern behavior, where the smaller slopes are directed towards the equatorial region, so that the north and south poles move loose material in direction to this region 
\citep{Scheeres2012,Scheeres2016}.
A behavior contrary to the one mapped on the surface of (16) Psyche, since the slope values are sensitive to the size, density, and rotational speed of the body.

It is worth noting that the mapping of the tilt and slope angles across the asteroid surface may differ considerably from one another, as illustrated in Figs. \ref{fig:tilt} and \ref{fig:slope}, 
since the tilt angle is purely geometric, and the variation of the slope angle depends on the rotation speed of the body.

The acting of the rotational dynamics of the asteroid influences the movement of material on the surface and may contribute to the alteration of the surface environment.
According to \citet{Scheeres2012}, the slope map of most bodies with fairly low slopes, whose values do not exceed $30^{\circ}$, can be indicative of a relaxed surface.
The low slopes across (16) Psyche are indicators of a relaxed surface, attesting that the migration of loose material on the surface probably occurred in the past \citep{Shepard2018}.

Taking into account the extreme values of density, the results are close to those of Fig. \ref{fig:slope}, also reflecting  high slopes in the southern hemisphere.
For the low density, the total variation of slopes is $21.1^{\circ}$ and $22.7^{\circ}$ for the high density. So, all the above analysis is also valid even for densities so different.


\section{Features of the dynamical environment}
\label{nearby environment}

In this section, we investigate the environment in the vicinity of (16) Psyche. The analysis of the zero-velocity curves allows evaluating regions where the movement of a particle will be 
allowed or not. Furthermore, they are a way of delimiting external equilibrium points in the gravitational field of (16) Psyche. Finally, the lower value of the energy evaluated at these 
equilibrium points establishes whether a particle may escape the surface of the asteroid, or whether its motion will be restricted to the environment close to the body.

\subsection{Zero-velocity curves}
\label{zero-velocity curves}

As defined in Subsection \ref{jacobi_constant}, the equation (\ref{eq: jacobi}) provides important information regarding the orbital motion in the environment around of (16) Psyche, since it delimits regions where a particle can move or 
not, according to the value of its Jacobi constant. Note that $\frac{1}{2}v^2 \geq 0$, then, the following inequality is true:
\begin{equation}
J - V(\pmb r) \geq 0.
\label{eq: inequality}
\end{equation}
Depending on the value of $J$ in the expression above, there will be regions where the trajectory of the particle will be restricted. The inequality is satisfied in the case of if 
$V(\pmb r) < J$, so that all space will be divided into regions allowed for the motion of the particle, at first without any limitation. In the case of $V(\pmb r) > J$, there will be 
regions in the space forbidden for particle motion, since the inequality is violated. If the arrangement of these forbidden regions demarcates space in several disjoint regions, 
the particle will be confined, since the trajectory between these regions is strictly prohibited. Consequently, $V(\pmb r) = J$ dictates the limit of these regions, denominated as 
zero-velocity surface, or zero-velocity curve in the planes. \citet{Scheeres1994} presents a more general and complete study about zero-velocity surfaces around the uniformly rotating 
triaxial ellipsoids.

\begin{figure*}
\begin{center}
\includegraphics[trim = 0mm 50mm 0mm 0mm, clip, width=\linewidth]{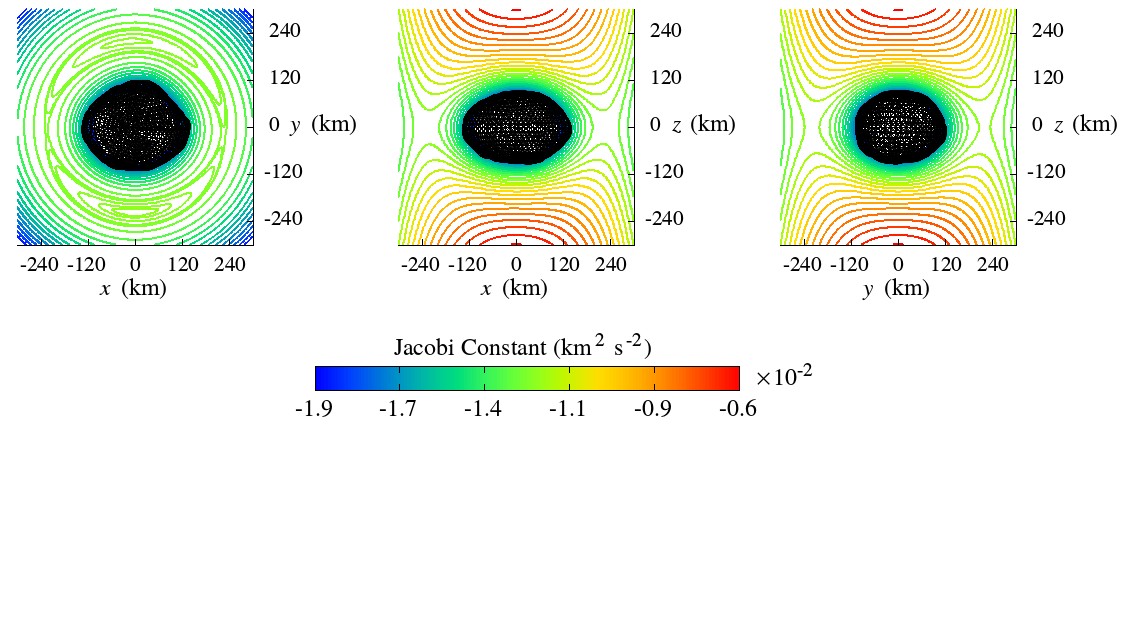}
\end{center}
\caption{Contourplots of zero-velocity curves of (16) Psyche in the $xoy$, $xoz$ and $yoz$ planes, respectively, for nominal density. The colour code provides the value of the Jacobi constant in units of km$^2$ s$^{-2}$.}
\label{fig: zero-velocity curves}
\end{figure*}

Fig. \ref{fig: zero-velocity curves} shows the contourplots of zero-velocity curves corresponding to the (16) Psyche shape model onto the $z = 0$, $y = 0$ and $x = 0$ planes, according to 
the intensity of the Jacobi constant. Zero-velocity curves are also useful for determining the location of equilibrium points. At these points the curves intersect or even 
close in upon themselves, to critical values of $J$. 

\subsection{Equilibrium points}
\label{points}

Several studies have computed four equilibrium points around of uniformly rotating small bodies, such as the asteroids (1620) Geographos, (4769) Castalia, and (216) Kleopatra 
\citep{Jiang2014, Wang2014}. These points are commonly referred to as $E_1$, $E_2$, $E_3$, and $E_4$. However, the number of equilibrium points can change according to the body 
considered. \citet{Wang2016} analyzed the gravitational field of (101955) Bennu and located eight external points near the surface of the body.

The equilibrium points are points at which there is a balance of forces acting on these positions, and to determine its location the condition imposed by equation (\ref{eq: condition}) 
must be satisfied
\citep{Jiang2014}:
\begin{equation}
\nabla V(\pmb r)=0.
\label{eq: condition}
 \end{equation}
The resolution of this equation in principle does not provide an exact amount of solutions since the irregular shape and rotation of the body influence the total equilibrium points 
present in a gravitational field \citep{scheeres1996}. 

\begin{figure}
\begin{center}
\includegraphics*[width=\columnwidth]{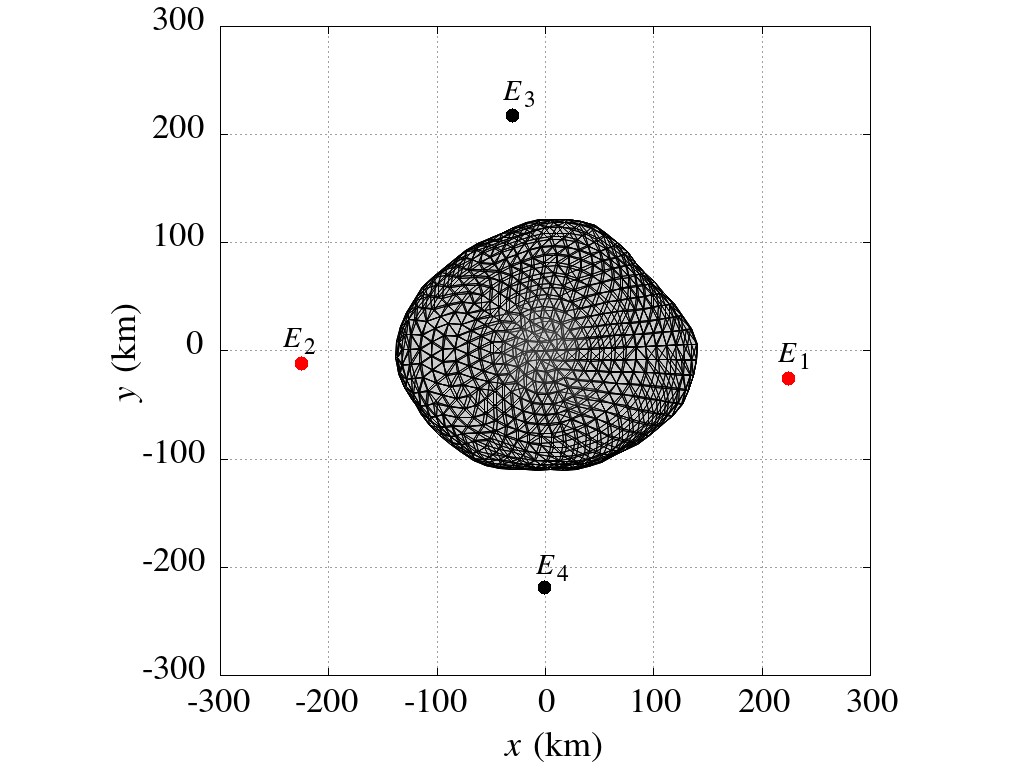}
\end{center}
\caption{Location of the asteroid equilibrium points (16) Psyche in the $xoy$ plane, for nominal density (4.5 g cm$^{-3}$). Black dots are linearly stable while red dots are unstable 
equilibrium points.}
\label{fig: points}
\end{figure}

As already mentioned in Subsection \ref{zero-velocity curves}, we can use the zero-velocity curves of (16) Psyche (Fig. \ref{fig: zero-velocity curves}) as an extra resource that facilitates the 
identification of the number of equilibrium points. Table \ref{table: value points with different densities} specifies the location of the external equilibrium points of (16) Psyche and Fig. 
\ref{fig: points} illustrates the arrangement of these points projected in the equatorial plane ($z = 0$), for nominal density. Note that the four points of equilibrium external to the body 
are roughly located close to the equatorial plane, and there is an almost symmetry about the $x = 0$ and $y = 0$ axes, despite the irregular shape of (16) Psyche. 

The linear stability and topological classification of the equilibrium points in the gravity field of more than 20 small celestial bodies were investigated in \citet{Wang2014}, according to the 
classification established in \citet{Jiang2014}. We apply a linearization method to examine the stability of the equilibrium points of (16) Psyche, and using the eigenvalues of the 
characteristic equation, we classify these points based on \citet{Jiang2014}. The characteristic equation produces six eigenvalues, and the topological classification of each equilibrium 
point is established according to the disposition of these eigenvalues in the complex plane. The analysis of the eigenvalues of table \ref{table: eigenvalues}, showed that the points 
$E_3$ and $E_4$ are classified topologically in Case 1 (each equilibrium point has three pairs of imaginary eigenvalues). Consequently, $E_3$ and $E_4$ are linearly stable. 
The eigenvalues of points $E_1$ and $E_2$ fit into Case 2 (there are two pairs of imaginary eigenvalues and one pair of real eigenvalues for the equilibrium point) and are classified as 
unstable points.

However, the location and topological classification of the equilibrium points in the gravitational field of an asteroid can undergo changes as the variation of a 
given parameter occurs, such as rotation speed and density. \citet{Jiang2015b} considered the rotation speed as a parameter that could be modified and investigated the behavior
of the equilibrium points in the gravitational field of the asteroid (216) Kleopatra. They observed that as the speed of rotation increased the number of equilibrium 
points decreased from 7 to 5 to 3 to 1. It happened that some points collided and annihilate each other until only one point inside the object remained.
In this way, some types of bifurcations are possible to appear at the moment when the coalescence occurs, and the points disappear. In addition, the topological classification of 
the points changes. A detailed study of the types of bifurcations is discussed in \citet{Jiang2015b}.

\begin{table}
\centering
\caption{Location of the equilibrium points of (16) Psyche considering the nominal (4.5 g cm$^{-3}$) and extreme density values.}
\label{table: value points with different densities} 
\scalefont{0.9}
\begin{tabular}{ccccc}
\hline\hline
   Equilibrium &     Density   & $x$  & $y$  & $z$ \\   
      point    & (g cm$^{-3}$) & (km) & (km) & (km)\\
\hline
\hline
 &3.1& 226.2786& $-27.2245$&$-0.5938$\\
$E_1$ &4.5& 224.2672& $-25.6407$&$-0.3685$\\
 &7.6& 222.3988&$-22.8875$& $-0.1967$\\
\hline
 &3.1& $-227.8522$& $-11.5003$&$-0.3783$\\
$E_2$ &4.5& $-225.6746$& $-11.5980$&$-0.2625$\\
 &7.6& $-223.4892$&$-11.3641$& $-0.1570$\\
\hline
 &3.1& $-40.3533$& $215.8981$&$-0.5922$\\
$E_3$ &4.5& $-30.6626$& $217.2641$& $-0.3628$\\
 &7.6& $-20.3769$&218.2330& $-0.1823$\\
\hline
 &3.1& $-0.3671$& $-217.9651$&$-0.0400$\\
$E_4$ &4.5& $-0.6472$&$-218.1959$& $-0.0399$\\
  &7.6&$-0.9801$&$-218.4222$&$-0.0306$\\
\hline
\end{tabular}
\end{table}

\begin{table}
\begin{center}
\caption{Eigenvalues ($\lambda_n \times10^{-3}$) of equilibrium points in the potential of (16) Psyche considering the nominal (4.5 g cm$^{-3}$) and extreme density values ($\rho$, in g cm$^{-3}$).}
\vspace{0.1mm}
\label{table: eigenvalues} 
\scalefont{0.9}
\begin{tabular}{crrrrr}
\hline\hline
   $\lambda_n$      &   $\rho$   & $E_1$ & $E_2$ & $E_3$ & $E_4$ \\ 
\hline
\hline
            & 3.1 & $0.4520i$& $0.4509i$& $0.4332i$& $0.4277i$\\
$\lambda_1$ & 4.5 & $0.4424i$& $0.4426i$& $0.4289i$& $0.4254i$\\
            & 7.6 & $0.4334i$& $0.4343i$& $0.4248i$& $0.4228i$\\
\hline
            & 3.1 &$-0.4520i$&$-0.4509i$&$-0.4332i$&$-0.4277i$\\
$\lambda_2$ & 4.5 &$-0.4424i$&$-0.4426i$&$-0.4289i$&$-0.4254i$\\
            & 7.6 &$-0.4334i$&$-0.4343i$&$-0.4248i$&$-0.4228i$\\
\hline
            & 3.1 & $0.4468i$& $0.4461i$& $0.3466i$&$-0.0556+0.2909i$\\
$\lambda_3$ & 4.5 & $0.4384i$& $0.4380i$& $0.3707i$&$0.3323i$\\
            & 7.6 & $0.4302i$& $0.4305i$& $0.3865i$&$0.3685i$\\
\hline
            & 3.1 &$-0.4468i$&$-0.4461i$&$-0.3466i$&$-0.0556-0.2909i$\\
$\lambda_4$ & 4.5 &$-0.4384i$&$-0.4380i$&$-0.3707i$&$-0.3323i$\\
            & 7.6 &$-0.4302i$&$-0.4305i$&$-0.3865i$&$-0.3685i$\\
\hline
            & 3.1 &$-0.2406$ &$-0.2374$ & $0.1954i$& $0.0556+0.2909i$\\
$\lambda_5$ & 4.5 &$-0.2048$ &$-0.2044$ & $0.1567i$& $0.2336i$\\
            & 7.6 &$-0.1648$ &$-0.1669$ & $0.1274i$& $0.1774i$\\
\hline
            & 3.1 & $0.2406$ & $0.2374$ &$-0.1954i$& $0.0556-0.2909i$\\
$\lambda_6$ & 4.5 & $0.2048$ & $0.2044$ &$-0.1567i$&$-0.2336i$\\
            & 7.6 & $0.1648$ & $0.1669$ &$-0.1274i$&$-0.1774i$\\
\hline
\end{tabular}
\end{center}
\end{table}

Thus, with the intention of ascertaining possible changes of the (16) Psyche equilibrium points as well as the topological structure of them, we chose the variation of the density parameter.
Our choice is based on the fact that this parameter is still uncertain, since previous works point to a large range for density \citep{Shepard2017, Drummond2018, Viikinkoski2018}.
We analyzed the location and linear stability of the equilibrium points for several values of densities in a range between 3.1 and 7.6 g cm$^{-3}$ (the extreme values).
Table \ref{table: value points with different densities} also identifies the location of the equilibrium points considering the extreme values of density.
Note that even considering extreme density values, the coordinates of the equilibrium points are very similar.
This is due to the fact that we are preserving the mass of (16) Psyche, and as density changes, we have a new volume, and therefore, a new size of the object. However, the gravitational 
force exerted is practically the same in the region of the equilibrium points, since this depends on the mass distribution of the asteroid.
The analysis of the eigenvalues (Table \ref{table: eigenvalues}) showed that, even with values of different densities, the topological classification of all the points is preserved, 
with the exception of point $E_4$. Following the topological classification presented in \citet{Jiang2014}, for the inferior extreme value of density (3.1 g cm$^{-3}$) the point $E_4$ 
belongs to Case 5 (there are a pair of imaginary eigenvalues and two pairs of conjugate complex eigenvalues to the equilibrium point).

In order to verify how large is the size of the stability regions around the four equilibrium points, considering the nominal density, we perform another set of simulations where we 
distribute massless particles in a region containing these points.
We then investigated how these particles initially placed in the vicinity of (16) Psyche and close to the external equilibrium points evolved under its 
gravitational potential, and to do so, the polyhedra shape used in the previous sessions was replaced by a mass concentrations (MASCONS) model \citep{Geissler1996}. Even though it is precise, 
the polyhedra model is not suitable to integrate the orbits due to the time necessary to evaluate the potential each time step. The mascons model circumvent the need of several calculations 
that raise from the integral formalism, but keeps good accuracy when compared with the polyhedra technique and gives a more precise representation for the potential close to the body 
surface when compared to a spherical harmonics representation \citep{WernerScheeres1996, Rossi1999}. Based on the method presented in \citet{Geissler1996}, an equally spaced mesh is created 
encompassing the entire asteroid. From this mesh, we select the points that lie within the asteroid volume, and to each point, a mass is attributed in such a manner that the sum corresponds to the
total mass of (16) Psyche. Therefore, the sum of the gravitational potential from each mass point is the gravitational potential from Psyche's irregular shape, as given by 
\citep{BorderesMotta2018}
\begin{equation}
U = U(x,y,z) = \sum^{N}_{i=0} \frac{G \ m}{r_{i}},
\label{eq: potential mascons}
 \end{equation}
where $N$ represents the number of mascons confined in the volume of (16) Psyche, $r_{i}$ defines the distance of the orbiting particle in relation to each mascon, whose mass is given by 
$m=\frac{M}{N}$. In our model we considered a $4.6$ km spaced grid, resulting in a model for (16) Psyche composed by $N=62422$ mascons of $m=4.3734\times10^{14}$ kg each.
For this number of mascons, we compute the relative error of gravitational potential between mascon and polyhedron model. Over the surface of the body, the average of the relative error 
is $\sim$0.08\% and the standard deviation is $\sim$0.06\%. Therefore, the mascon model represents very well the potential of the polyhedron model.

The initial conditions constitute an ensemble of 15 thousand massless particles distributed in 30 equally spaced layers with $195 < r < 255$ km, ranging from some tens of kilometers above the 
equivalent radius of the body to a few kilometers beyond the location of the equilibrium points. In each layer, 500 particles were placed initially in circular orbits and in the equatorial plane.
To create a flat cloud, random values from $0^{\circ}$ to $360^{\circ}$ were attributed to the  mean anomaly.
The equations of motion we integrated with the Bulirsch-Stoer algorithm in a modified version of the Mercury package \citep{chambers1999} that includes the mascons model. We followed the orbits 
for a timespan of 2 years ($\sim$ 4 thousand spin periods of (16) Psyche).

\begin{figure}
\begin{center}
\includegraphics*[width=0.98\columnwidth]{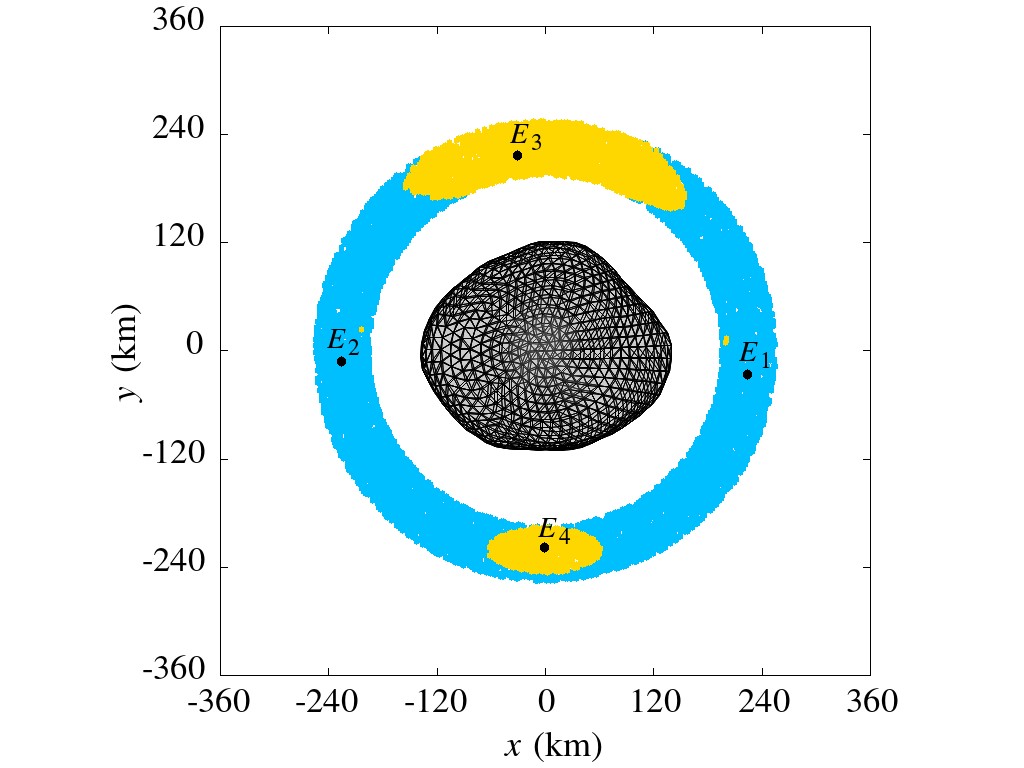}
\end{center}
\caption{Blue represents the positions in the $xoy$ plane of 15 thousand particles randomly distributed around (16) Psyche at the initial time.
   Yellow represents the initial positions of the particles that remained after two years of integration. Black points represent the position of the equilibrium points 
   for the nominal density.}
\label{fig: particles}
\end{figure}

Let us assume that a stable region is one where there are particles that survived until the end of the simulation, were not ejected and did not collide with the central body. 
Fig. \ref{fig: particles} shows the initial position of the particles of the initial ensemble (blue points) and those that survived until the end of the simulation (yellow dots).
The results remarkably show that the size of the stable regions around the points $E_3$ and $E_4$ are asymmetric.
The region around point $E_3$ has an angular amplitude of $\sim$ $81^{\circ}$, while for $E_4$ it is only $\sim$ $30^{\circ}$.
This discrepancy in the sizes of the stability regions around points $E_3$ and $E_4$ is certainly due to the asymmetric shape of (16) Psyche.

\begin{figure*}
\begin{center}
\includegraphics[height=9cm, width=13cm]{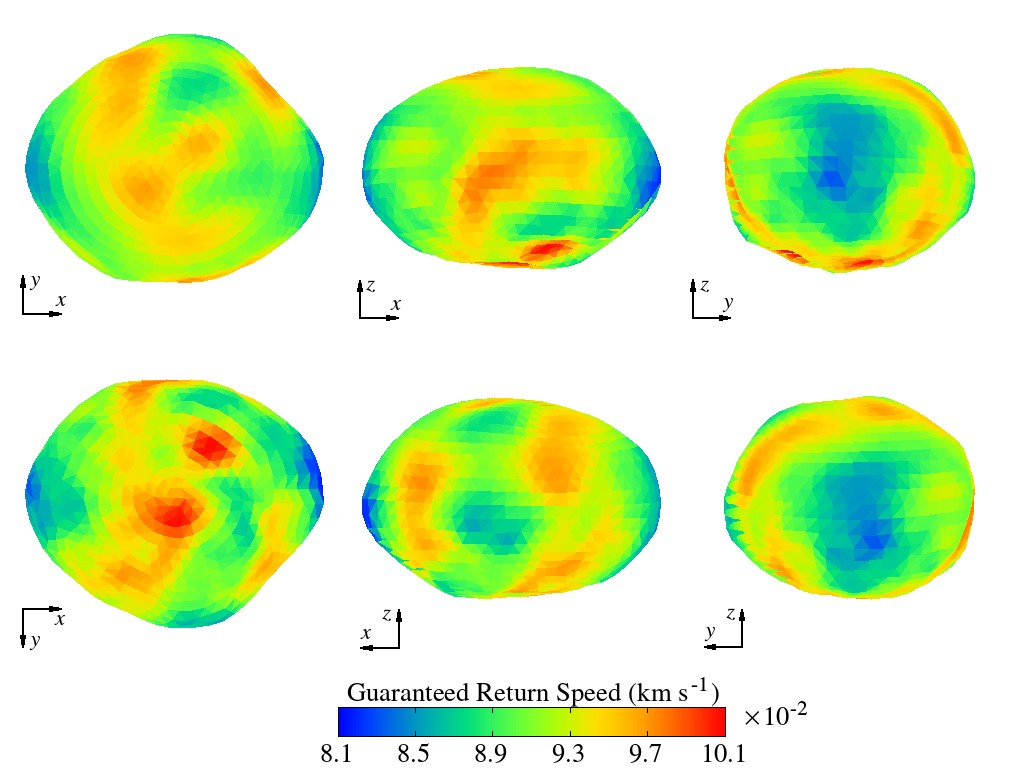}
\end{center}
\caption{Guaranteed return speed $v_{\rm ret}$ mapped across the surface of (16) Psyche considering the nominal density, under different views.}
\label{fig:return_speed}
\end{figure*}

\subsection{Guaranteed return speed}
\label{ret_speed}
(16) Psyche, like the vast majority of asteroids, has four synchronous orbits (external equilibrium points) in its vicinity, as shown in Subsection \ref{points}. Obviously, some bodies do not 
adhere to this prescription, as is the case with the asteroid (101955) Bennu, holder of eight external equilibrium points \citep{Scheeres2016}.

Considering a system that is fixed and rotates with (16) Psyche, the geopotential computed at these points of equilibrium produces an energy able of defining limits in the movement of a 
particle near the asteroid. Such limit is imposed by the zero-velocity surface (or zero-velocity curves in the coordinate planes) generated by the smaller value of the Jacobi constant ($J^*$), 
among those computed by means of the geopotential measured in the synchronous orbits \citep{Scheeres2012}. This surface surrounds the entire asteroid in three-dimensional space and its minimal 
energy makes it possible to analyze the lower limit of the guaranteed return speed, a velocity evaluated on the surface of the body as:
\begin{equation}
\label{eq: return speed}
v_{\rm ret} = \sqrt{-2(V(\pmb r)-J^*)},
\end{equation}
whose value is set to zero if the geopotential across the surface of the body exceeds $J^*$. Recalling that for (16) Psyche we take into account the value of the geopotential computed in the 
barycentre of each triangular face of the polyhedral model of the body.

Thus, a particle at a location $\pmb r$ on the surface of (16) Psyche will have an energy $V(\pmb r)$ and a return speed $v_{\rm ret}$. If this energy is greater than $J^*$, then $v_{\rm ret}=0$ 
and it is possible that this particle can escape from the environment near the asteroid, even though it is on the surface of (16) Psyche. Otherwise, if this energy is less than $J^*$, and considering that the particle is positioned within the zero-velocity curve that surrounds the body, there is insufficient energy for it to escape from the system. 
So, $v_{\rm ret}$ will define a lower bound for the velocity across the surface, below which the particle will be conditioned to have its motion restricted to the neighborhood of (16) Psyche.

Fig. \ref{fig:return_speed} illustrates the behavior of the guaranteed return speed across the surface of (16) Psyche, considering the nominal density. The value of $J^*$ is given by 
$-1.2697\times 10^{-2}$ km$^2$ s$^{-2}$ and is associated with the equilibrium point $E_2$. Observe that the minimum values of the return speed are at the ends of the equator, while the maximum 
values are at the poles.
Moreover, according to Fig. \ref{fig: view_speeds}, a particle that starts its trajectory from the highest point in the geopotential toward the lowest point requires a 
speed of up to 0.06 km s$^{-1}$ to boost such a movement. Value below the lower limit of the return speed established in the polar regions (0.101 km s$^{-1}$), where the lowest point in the 
geopotential is concentrated. Thus, it is impossible the movement of particles outside the zero-velocity curve that encompasses the asteroid since their energies are smaller than $J^*$.
Consequently, a particle within this zero-velocity curve can not leave the close proximity of (16) Psyche.

\section{Mapping of impacts across the surface}
\label{collision}

\begin{figure*}
\begin{center}
\subfloat{\includegraphics*[trim = 0mm 50mm 0mm 45mm, width=\linewidth]{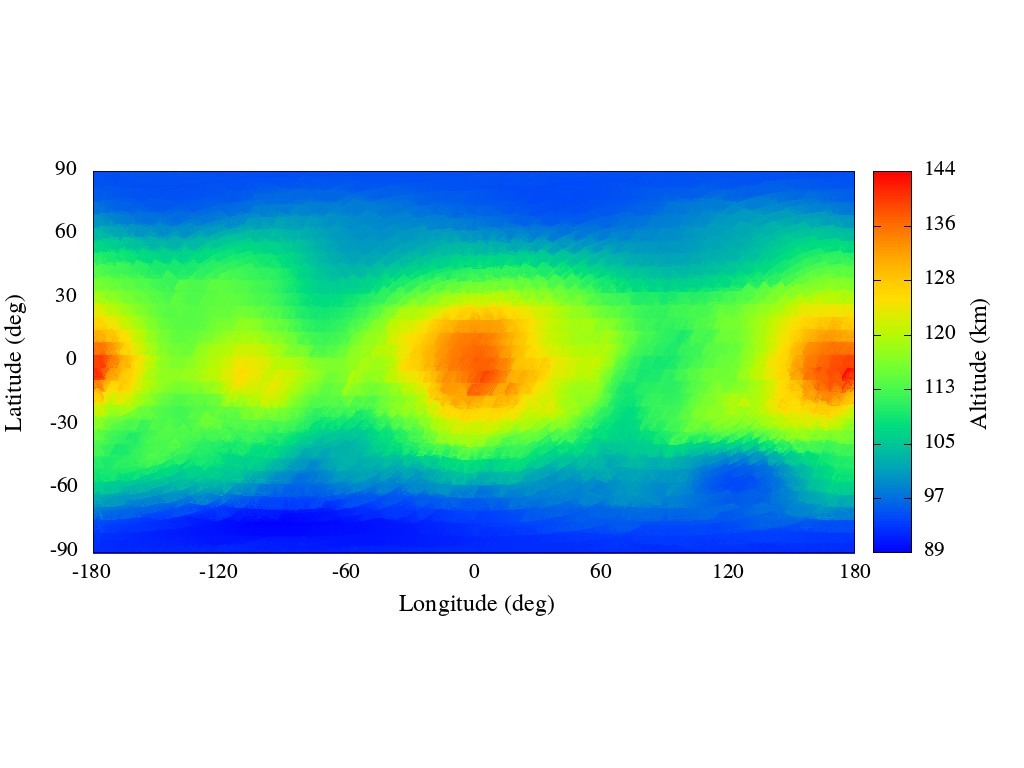}}\\
\subfloat{\includegraphics*[trim = 0mm 45mm 0mm 42.5mm, width=\linewidth]{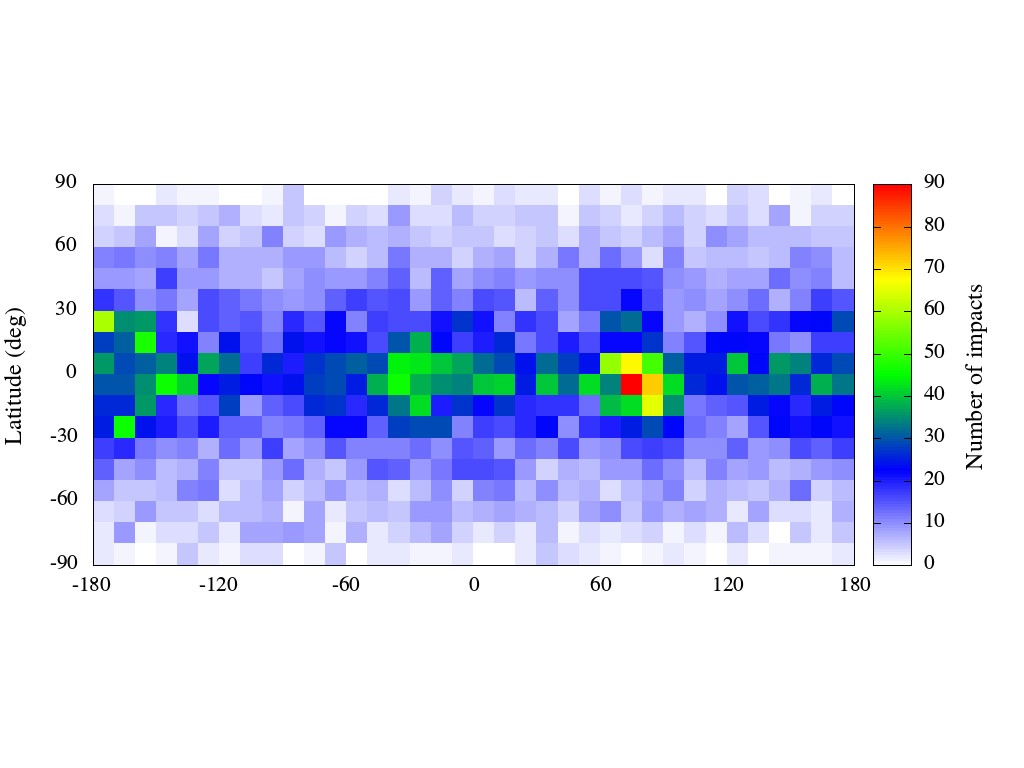}}\\
\end{center}
\caption{Top: Altitude of astroid (16) Psyche in longitude and latitude. The color code provides the distance from the center of mass of the body, measured in km.
Bottom: Number of fall of particles across the surface of (16) Psyche mapped in a $10^\circ \times 10^\circ$ grid in an equirectangular projection.}
\label{fig: impacts}
\end{figure*}

Considering that throughout their history, planetary satellites frequently undergo collisions with interplanetary dust particles \citep{Kruger2000, Kruger2003, Krivov2003, Winter2018}, 
generating ejecta that can initially orbit around the satellite. The same is also expected to occur with minor celestial bodies as the asteroid (16) Psyche. Thus, in this subsection, 
we are interested in mapping the fall of the ejecta particles on the surface of (16) Psyche. So, let us investigate whether there is a preference for certain regions and how altitude can 
influence the fall of these particles.

An ensemble of 10 thousand test particles was distributed in 40 equally spaced layers with $126.5 < r < 241.5$ km, ranging from few kilometers above the equivalent radius of the body until 
comprising the equilibrium points. At each layer, 250 particles were initially in circular orbits, and the inclination was randomly chosen from $0^{\circ}$ to $90^{\circ}$. To create a 
cloud, random values from $0^{\circ}$  to $360^{\circ}$ were attributed to the longitude of the ascending node and the mean anomaly.

Here, again the equations of motion were integrated with the Bulirsch-Stoer algorithm considering a modified version of the Mercury package \citep{chambers1999} that includes the mascons model  
by the same arguments already described at the end of Subsection \ref{points}. We followed the orbit of each particle for a timespan of 1 year ($\sim$ 2 thousand spin periods of (16) Psyche). 
Every time that a particle collided with the asteroid, it was removed from the system and the impact position was registered.

A graph of the altitude measured from the centre of mass of (16) Psyche is shown in Fig. \ref{fig: impacts}-Top. The longitude was measured from a prime meridian defined by the direction of 
the axis of the smallest moment of inertia. While Fig. \ref{fig: impacts}-Bottom shows the collisions mapped in a $10^\circ \times 10^\circ$ grid in an equirectangular projection. In one spin period 
of (16) Psyche, about 82$\%$ of the particles had already fallen on its surface. Although the particles were distributed in a cloud around the asteroid and almost all the surface was impacted, 
one can note the collisions happened primarily around the equatorial region in a latitude range around $\pm 30^\circ$, which corresponds to the region with the highest altitudes as discussed 
in subsection \ref{geometric topography}. When comparing Fig. \ref{fig: impacts}-Bottom with Fig. \ref{fig: impacts}-Top, we note that the places with a larger number of collisions (green and red in 
Fig. \ref{fig: impacts}-Bottom) correspond to lower altitudes, mainly the one located at longitude $70^\circ$. This concentration is due to the gravitational pull experienced by the particle 
when flying close to the high altitude points located just to the left of the prime meridian (red region) that deviates the orbit towards the asteroid, eventually 
leading to a collision. The same also happens for the equatorial region with longitude between $-170^\circ$ and $-140^\circ$ (green dots in Fig. \ref{fig: impacts}-Bottom), which occurs to be just 
after the high altitude region shown in Fig. \ref{fig: impacts}-Top (red dots with longitudes between $160^\circ$ and $-170^\circ$).
The low number of collisions at high latitudes might be due to the somewhat ellipsoidal shape of the object, what makes the polar regions flattened.

The detection of material deposits in this region near the equator by a future mission may indicate that in the past, there was material orbiting (16) Psyche.


\section{Final Comments}
\label{conclusions}
This study provided the exploration of the dynamical environment around and on the surface of the asteroid (16) Psyche, the target of a future space mission.
Firstly, we exposed the shape model of the object and its physical properties, highlighting the discordance in the density value \citep{Shepard2017, Drummond2018, Viikinkoski2018}.
Then we define the gravitational potential using the polyhedra method \citep{WernerScheeres1996} and then present the geopotential computed on the surface of the body.

Next, we presented the topographic features on the surface of (16) Psyche utilizing the geopotential. We compute the geopotential topography, potential speeds, 
surface accelerations, tilts, and slopes. These characteristics took into consideration the nominal density and the extreme values presented in \citet{Shepard2017}.
The altitude variation considering the body geometry and the effective altitude variation considering the geopotential were mapped across the surface of (16) Psyche.
The regions on the asteroid surface that comprise the lowest and highest points in the geopotential were located. And an analysis of the tilt angles and slopes were made in relation to 
these regions. We present a purely geometric amount, tilt angle, which provides how the surface is oriented in the body system. The importance of this angle is due to the fact that in case a
probe makes contact normal to the surface of (16) Psyche, it is necessary the orientation in relation to that direction.
The mapping of slopes provides information on the direction of the movement of loose material on the surface of the object.
We emphasize the significant difference in the distribution of slopes between the northern and southern hemispheres. The slopes close to zero are concentrated in the northern hemisphere, which 
characterizes it as a more ``relaxed'' region.

With the support of zero velocity curves, it was possible to determine the number of equilibrium points in the gravitational field of this object, considering the nominal density. 
We found that two of the four external equilibrium points are linearly stable ($E_3$ and $E_4$).
A discussion of how the density affects the location and linear stability of these points was given, taking into account the variation of this parameter in a range between 3.1 and 7.6 
g cm$^{-3}$. In addition, the location and linear stability of the equilibrium points aid in understanding the environment around the object.
We used a modified version of the Mercury package \citep{chambers1999} to integrate for two years particles without mass in the vicinity of (16) Psyche, and the result showed that the 
stable region around the point $E_3$ is almost three times larger in angular amplitude than the region around the point $E_4$.

Besides, we have mapped the impact of particles across the surface of (16) Psyche, noting that collisions occur preferentially in the equatorial region of the body, precisely between high altitudes.

This study can be useful for motivating and assisting in the observation plans for the Psyche mission. All the analysis presented will be able to measure our ability to investigate the 
geophysical environment of asteroids, using a tool the Earth-based observations. Moreover, the Psyche mission may test or validate this study and point out characteristics that we do not 
cover.


\section*{Acknowledgements}

This research was financed in part by the Coordena\c{c}\~ao de Aperfei\c{c}oamento de Pessoal de N\'ivel Superior - Brasil (CAPES) - Finance Code 001, FAPESP (Proc. 2016/24561-0), 
CNPQ (Proc. 305737/2015-5) and CNPQ (Proc. 305210/2018-1). 

We are grateful to the reviewer for the suggestions that have contributed to a substantial improvement of the paper.

\section*{ORCID iDs}
T. S. Moura \orcidicon{0000-0002-3991-8738} \href{https://orcid.org/0000-0002-3991-8738}{https://orcid.org/0000-0002-3991-8738}\\
O. C. Winter \orcidicon{0000-0002-4901-3289} \href{https://orcid.org/0000-0002-4901-3289}{https://orcid.org/0000-0002-4901-3289}\\
A. Amarante \orcidicon{0000-0002-9448-141X} \href{https://orcid.org/0000-0002-9448-141X}{https://orcid.org/0000-0002-9448-141X}\\
R. Sfair \orcidicon{0000-0002-4939-013X} \href{https://orcid.org/0000-0002-4939-013X}{https://orcid.org/0000-0002-4939-013X}\\
G. Borderes-Motta \orcidicon{0000-0002-4680-8414} \href{https://orcid.org/0000-0002-4680-8414}{https://orcid.org/0000-0002-4680-8414}\\
G. Valvano \orcidicon{0000-0002-7905-1788} \href{https://orcid.org/0000-0002-7905-1788}{https://orcid.org/0000-0002-7905-1788}




\bibliographystyle{mnras}
\bibliography{bilbli} 





\bsp	
\label{lastpage}
\end{document}